\documentclass[11pt,twoside,a4paper]{article}
\usepackage[english]{babel}
\usepackage{amsmath,amsfonts,amssymb,bm}\interdisplaylinepenalty=2500
\usepackage{longtable}

 \mathchardef\epsilon="0122   \mathchardef\varepsilon="010F
 \mathchardef\theta="123      \mathchardef\vartheta="0112
 \mathchardef\rho="125        \mathchardef\varrho="011A
 \mathchardef\phi="127        \mathchardef\varphi="011E
\ifx\undefined\degrees\def\degrees{\ensuremath{^{\circ}}}\fi
\ifx\undefined\celsius\def\celsius{\ensuremath{^{\circ}\mathrm{C}}}\fi
\ifx\undefined\unit\def\unit#1{\ensuremath{\mathrm{\,#1}}}\fi
\ifx\undefined\micro\def\micro{\ensuremath{\mu}}\fi
\ifx\undefined\sups\def\sups#1{\ensuremath{^{\mathrm{#1}}}}\fi
\ifx\undefined\subs\def\subs#1{\ensuremath{_{\mathrm{#1}}}}\fi
\ifx\undefined\ohm\def\ohm{\ensuremath{\mathrm{\Omega}}}\fi
\def\req#1{(\ref{#1})}
\def\dr{\vrule width 0pt height 1.2ex depth 1.2ex}
\def\ur{\vrule width 0pt height 2.5ex depth 0.0ex}

\ifx\undefined\p\def\p#1{\ensuremath{_\textsc{#1}}}\fi
\def\shrinklist{\itemsep-0.3ex}

\newtheorem{rremark}{Remark}%[chapter]

\newtheorem{eexample}{Example}%[chapter]
\newenvironment{example}{%
	\begin{eexample}\normalfont}{\mbox{}%
	\hfill$\blacktriangle$\end{eexample}}
\newtheorem{pproperty}{Property}%[chapter]

\newtheorem{ddefinition}{\noindent\bfseries\itshape}

\usepackage{graphicx,color}   % for including graphics
\usepackage{rotating}

\def\LARGEscale{0.917}
\def\normalscale{0.707}
\def\smallscale{0.648}
\title{Tutorial on the double balanced mixer}
\author{Enrico Rubiola\\
\small web page \texttt{http://rubiola.org}
\\[4em]\includegraphics[width=0.35\textwidth]{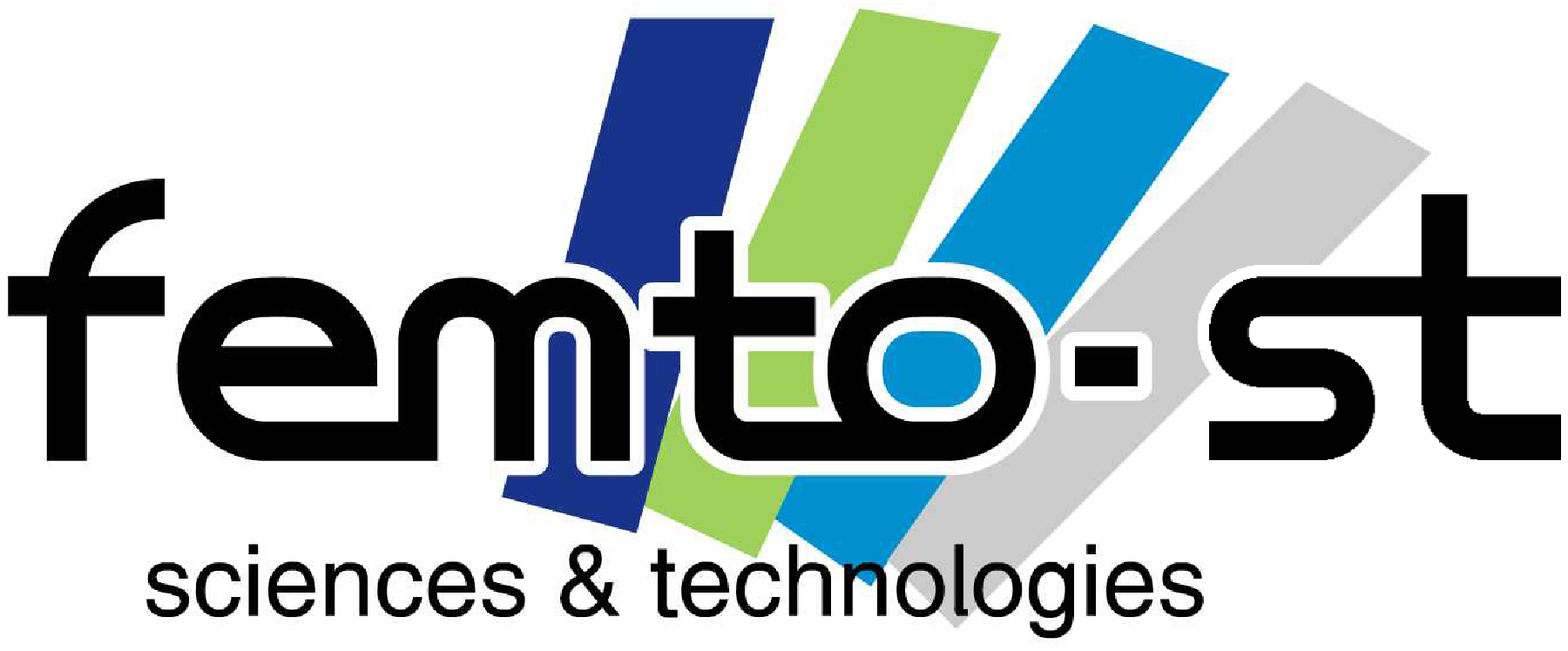}\\[0.5em]
\small FEMTO-ST Institute\\[-0.5ex]
\small CNRS and Universit\'e de Franche Comt\'e, 
\small Besan\c{c}on, France\\[1.5em]}
\date{\small\today}
\begin{document}
\maketitle
%===========================================
\begin{abstract}
%===========================================
Smart use of mixers is a relevant issue in radio engineering and in instrumentation design, and of paramount importance in phase noise metrology.  However simple the mixer seems, every time I try to explain to a colleague what it does, something goes wrong.  One difficulty is that actual mixers operate in a wide range of power (150 dB or more) and frequency (up to 3 decades).  Another difficulty is that the mixer works as a multiplier in the time-domain, which is necessary to convert frequencies.  A further difficulty is the interaction with external circuits, the input sources and the load.  Yet far the biggest difficulty is that designing with mixers requires a deep comprehension of the whole circuit at \emph{system} level and at a \emph{component} level.
As the electronic-component approach is well explained in a number of references, this tutorial emphasizes the system approach, aiming to provide \emph{wisdom} and \emph{insight} on mixes.
\end{abstract}

\clearpage
\begin{center}\begin{small}
\addcontentsline{toc}{section}{Most used symbols}
\begin{longtable}{ll}\hline
	%\begin{tabular}{ll}\hline
\multicolumn{2}{l}{\textbf{\large\rule[-1ex]{0pt}{3.5ex}Most used symbols}}\\\hline
\rule{0pt}{2.5ex}%
$A(t)$			& slow-varying (baseband) amplitude\\
$h_{lp}$, $h_{bp}$	& impulse response of lowpass and bandpass filters\\
$h$, $k$, $n$, $p$, $q$& integer numbers\\
$i(t)$, $I$		& current\\
I (goes with Q)	& in-phase in/out (of a two-phase mixer/modulator)\\  
IF				& intermediate frequency\\
$j$				& imaginary unit, $j^2=-1$\\
$\ell$		& mixer voltage loss, $1/\ell^2=P_i/P_o$\\
LO				& local oscillator\\
$P$			& power\\
$P_i$, $P_o$	& power, input and output power \\
$P_p$, $P_S$	& LO (pump) power and internal LO saturation power\\
Q (goes with i)	& quadrature in/out (of a two-phase mixer/modulator)\\  
$R$			& resistance\\
$R_0$			& characteristic resistance (by default, $R_0=50$ \ohm)\\
$R_G$			& source resistance (Th\'evenin or Norton model) \\
$U$			& dimensional constant, $U=1$\unit{V}\\
$v(t)$, $V$		& voltage\\
$v'$, $v''$		& real and imaginary, or in-phase and quadrature part\\
$v_i(t)$, $v_o(t)$	& input (RF) voltage, and output (IF) voltage\\
$v_p(t)$		& LO (pump) signal \\
$v_l(t)$, $V_L$	& internal LO signal\\
$V_O$		& saturated output voltage\\
$V_S$		& satureted level of the internal LO signal $v_l(t)$\\
$x(t)$			& real (in-phase) part of a RF signal\\
$y(t)$			& imaginary (quadrature) part of a RF signal\\
$\varphi$, $\varphi(t)$	& static (or quasistatic) phase\\
$\phi(t)$		& random phase\\
$\omega$, $f$	& angular frequency, frequency\\  
$\omega_i$, $\omega_l$ 
                   & input (RF) and  pump (LO) angular frequency\\
$\omega_b$, $\omega_s$	& beat and sideband angular frequency\\
\multicolumn{2}{l}{\small\bfseries note: \boldmath $\omega$ is used as a shorthand for $2\pi f$}%
\rule[-1ex]{0pt}{0ex}\\\hline
\multicolumn{2}{l}{\textbf{\rule[-1ex]{0pt}{3.5ex}Most used subscripts}}\\\hline
$b$				& beat, as in $|\omega_s-\omega_i|=\omega_b$\\  
$i$, $I$			& input\\  
$l$, $L$		& local oscillator (internal signal)\\  
$o$, $O$		& output\\  
$p$, $P$		& pump, local oscillator (at the input port)\\  
$s$				& sideband, as in $|\omega_s-\omega_i|=\omega_b$\\  
$S$			& saturated\\  
\multicolumn{2}{l}{\small\bfseries\boldmath note: in reverse modes, $i$ is still the input, and $o$ the output}\\\hline
\end{longtable}
\end{small}\end{center}
\clearpage\tableofcontents
\cleardoublepage

%===========================================
\section{Basics}
\begin{figure}[hb]
\centering%     % COMBINED LATEX-PS/PDF FIGURES
	\centering\scalebox{\normalscale}{%  %              % scale
	\input{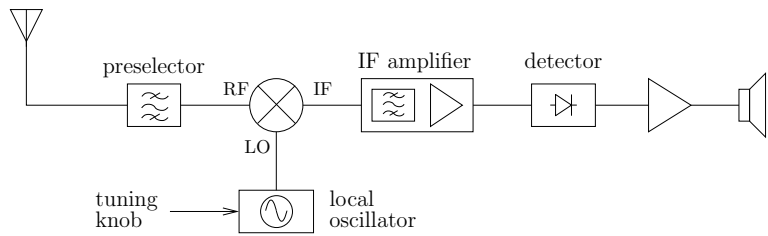}}
\caption{Superheterodyne receiver.}
\label{fig:mix-superhet}
\end{figure}

%===========================================
It is first to be understood that the mixer is \emph{mainly intended}, and \emph{mainly documented}, as the frequency converter of the superheterodyne receiver (Fig.~\ref{fig:mix-superhet}).  
The port names, LO (local oscillator, or \emph{pump}), RF (radio-frequency), and IF (intermediate frequency) are clearly inspired to this application.

\begin{figure}[ht]
\centering%     % COMBINED LATEX-PS/PDF FIGURES
	\centering\scalebox{\normalscale}{%  %              % scale
	\input{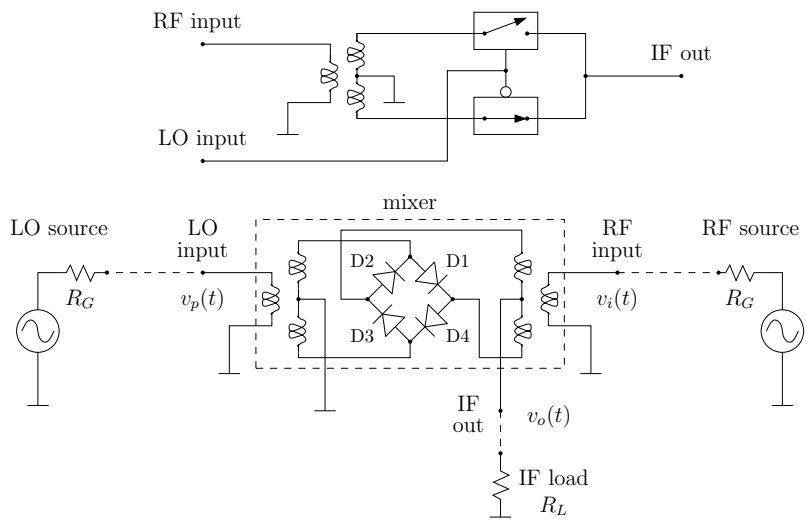}}
\caption{Double balanced mixer and its switch-network equivalent.}
\label{fig:mix-dbm}
\end{figure}

The basic scheme of a mixer is shown in Fig.~\ref{fig:mix-dbm}.  At microwave
frequencies a star configuration is often used, instead the diode ring.
Under the basic assumptions that $v_p(t)$ is large as compared to the diode threshold, and that $v_i(t)$ is small, the ring acts a switch.
During the positive half-period of $v_p(t)$ two diodes are reverse biased and the other two diodes are forward biased to saturation.  During the negative  half-period the roles are interchanged.  For the small RF signal, the diodes are open circuit when reverse biased, and small resistances when forward biased.  As a result, the IF signal $v_o(t)$ switches between $+v_i(t)$ and $-v_i(t)$ depending on the sign of $v_p(t)$.  This is equivalent to multiplying $v_i(t)$ by a square wave of amplitude $\pm1$ that takes the sign from $v_p(t)$.
In most practical cases, it is sufficient to describe the frequency conversion mechanism as the product between $v_i(t)$ and the first term of the Fourier expansion of the square wave.
More accurate models account for the higher-order Fourier terms, and for the dynamic resistance and capacitance of the diodes.

At the RF and LO sides, a balun is necessary in order to convert the unbalanced inputs into the balanced signals required for the ring to operate as a switch.  Conversely, no adapter is needed at the IF output, which is already unbalanced. 
In low-frequency mixers (from a few kHz to 2--3 GHz) the baluns are implemented with power iron tore transformers.  At higher frequencies, up to some tens of GHz, transformers are not available, for microstrip networks are the preferred balun types.  
The typical LO power is of 5--10 mW (7--10 dBm), whereas in some cases a power up to 1 W (30 dBm) is used for highest linearity.  The RF power should be at least 10 dB lower than the LO power.  The diodes are of the Schottky types, because of the low forward threshold and of the fast switching capability.  The characteristic impedance to which all ports should be terminated is $R_0=50$~\ohm, with rare exceptions.

The mixer can be used in a variety of modes, each with its ``personality'' and peculiarities, listed in Table~\ref{tab:mix:modes}, and detailed in the next Sections.  In short summary, the mixer is (almost) always used with the LO input saturated at the nominal power.  Then, the main parameters governing the behavior are:
\begin{description}
\item[Input power.] The input (RF) power is usually well below the saturation level, as in Figures \ref{fig:mix-superhet}--\ref{fig:mix-dbm}.  Yet, the input can be intentionally saturated. 
\item[Frequency degeneracy.] When the input (RF) and LO frequency overlap, the conversion product also overlap.  
\item[Interchanging the RF and IF ports.]  The difference is that the RF port is coupled in ac, while the IF port is often coupled in dc.
\end{description}
Additionally, the mixer is sometimes used in a \textbf{strange mode}, with both LO and RF inputs not saturated.

\def\clwa{6ex}\def\clwb{38ex}%normalsize
\begin{table}[t]
\centering
\caption{Operating modes of the double balanced mixer.\dr}
\label{tab:mix:modes}
\begin{tabular}{|c|c|cc|l|}\hline
&mode&\multicolumn{2}{c|}{\ur condition} & note \\
&\dr & frequency  & $P$ or $I$ & \\\cline{2-5}
%---------------------------------------------------------------- 
&LC  & $\nu_i\ne \nu_l$ & $P_i\ll P_S$ 
     & \parbox{\clwb}{\ur\emph{Linear} frequency \emph{Converter}.
       Typical of the superetherodyne radio receiver.\dr} 
       \\\cline{2-5}
&SD  & $\nu_i=\nu_l$ & $P_i\ll P_S$ 
     & \parbox{\clwb}{\ur\emph{Synchronous Detector}.  Used
       the lock-in amplifiers, in coherent receivers, and
       in bridge noise measurements.\dr}
       \\\cline{2-5}
&SC  & $\nu_i\ne\nu_l$ & $P_i\ge P_S$ 
     & \parbox{\clwb}{\ur\emph{Saturated} frequency \emph{Converter}.\\
       Mainly used in frequency synthesis.\dr}
       \\\cline{2-5} 
&DC  & \parbox{\clwa}{\centering$\nu_l{=}p\nu_0$\\[-0.3ex]
	$\nu_i{=}q\nu_0$\\\mbox{$p$, $q$ small integers}}
	&\raisebox{1.5ex}{$\smash{P_i\ge P_S}$} 
     & \parbox{\clwb}{%
     	\ur\emph{Degenerated} frequency \emph{Converter}.\\
       Only used in some cases of metrology and frequency 
       synthesis.\dr}
       \\\cline{2-5} 
\hspace*{1ex}\begin{rotate}{90}
	\hspace*{11ex}Normal Modes\end{rotate}
&PD  & $\nu_i=\nu_l$ & $P_i\ge P_S$
     & \parbox{\clwb}{\ur\emph{Phase Detector}.
       RF and LO signals are to be in quadrature.\dr}
       \\\hline
%---------------------------------------------------------------- 
&LM  & $\nu_i\approx0$ & $I_i\ll I_S$ 
     & \parbox{\clwb}{\ur\emph{Linear Modulator}, driven with 
       a near-dc input current $I_i(t)$.\dr} 
       \\\cline{2-5}
&RLC & $\nu_i\gg0$ & $P_i\ll P_S$ 
     & \parbox{\clwb}{\ur\emph{Reverse Linear Converter}, 
       driven with a narrowband signal at $\nu_i$.\dr} 
       \\\cline{2-5}
&DM  & $\nu_i\approx0$ & $P_i\ge P_S$
     & \parbox{\clwb}{\ur\emph{Digital Modulator}.
       Information is located close to dc.\dr}
       \\\cline{2-5} 
&RSC & $\nu_i\gg0$ & $P_i\ge P_S$
     & \parbox{\clwb}{\ur\emph{Reverse Saturated Converter}.
       Some cases of in frequency synthesis.\dr}
       \\\cline{2-5} 
\hspace*{1ex}\begin{rotate}{90}
\hspace*{6ex}Reverse Modes\end{rotate}
&RDC &  \parbox{8ex}{%
	\centering$\nu_l{=}p\nu_0$\\[-0.3ex]$\nu_i{=}q\nu_0$\\
	\mbox{$p$, $q$ small integers}}
	&\raisebox{1.5ex}{$\smash{P_i\ge P_S}$}
     & \parbox{\clwb}{\ur\emph{Reverse Degenerated Converter}.
       Similar to the DC mode, and only used in some odd cases.\dr}
       \\\hline 
\hspace*{1ex}\begin{rotate}{90}\hspace*{-3ex}Strange\end{rotate}
&AD &  $\nu_i=\nu_l$
                & \parbox{\clwa}{\centering$P_i{<}P_S$\\[0.5ex]$P_l{<}P_S$} 
     & \parbox{\clwb}{\ur\emph{Amplitude-modulation detector}.\\
       Scarce information.  Used at NIST for the measurement of AM noise.\dr}
       \\\hline 
\end{tabular}
\end{table}

%----------------------------------------------------------------
\subsection{Golden rules}
%----------------------------------------------------------------
\begin{enumerate}
\item First and foremost, check upon saturation at the LO port and identify the operating mode (Table \ref{tab:mix:modes}).

\item Generally, all ports should be reasonably impedance matched, otherwise reflected waves result in unpredictable behavior.

\item When reflected waves can be tolerated, for example at low frequencies or because of some external circuit, impedance plays another role.
In fact, the appropriate current flow is necessary for the diodes to switch.

\item In all cases, read carefully Sections \ref{ssec:mix:lc-mode} to \ref{ssec:mix:linearity}.
\end{enumerate}

%----------------------------------------------------------------
\subsection{Avoid damage}\label{sec:mix:safe-op}
%----------------------------------------------------------------
However trivial, avoid damage deserves a few words because the device can be pushed in a variety of non-standard operation modes, which increases the risk.
\begin{enumerate}
\item Damage results from excessive \emph{power}.  
Some confusion between maximum power for linear operation and the absolute maximum power to prevent damage is common in data sheets.

\item The nominal LO power (or range) refers to best performance in the linear conversion mode.  This value can be exceeded, while the absolute maximum power can not.

\item The maximum RF power is specified as the maximum power for linear operation.  When linearity is not needed this value can be exceeded, while the absolute maximum power can not.

\item Voltage driving may result in the destruction of the mixer for two reasons.  The diode $i=i(v)$ characteristics is exponential in $v$, for the current tend to exceed the maximum when the diode is driven by a voltage source.  The thin wires of the miniature transformers tend to blow up as a fuse if the current is excessive.

\item In the absence of more detailed information, the absolute maximum power specified for the LO port can be used as the \emph{total dissipated power}, regardless of where power enters.
  
\item The absolute maximum LO power can also be used to guess the \emph{maximum current through one diode}.   This may be useful in dc or degenerated modes, where power is not equally split between the diodes.
\end{enumerate}
Better than general rules, a misfortunate case occurred to me suggests to be careful about subtle details.  A \$\,3000 mixer used as a  phase detector died unexpectedly, without being overloaded with microwave power.  Further analysis showed that one rail of a dc supply failed, and because of this the bipolar operational amplifier (LT-1028) connected to the IF port sank a current from the input (20 mA?).

%=============================================
\section{Signal representations}\label{sec:mix:multiplication}
%=============================================
The simple sinusoidal signal takes the form
\begin{align}
\label{eqn:mix:simple-sinusoid}
v(t)=A_0\cos(\omega_0t+\varphi)~.
\end{align}
This signal has rms value $A_0/\sqrt2$ and phase $\varphi$.
An alternate form often encountered is 
\begin{align}
\label{eqn:mix:cos-sin-sinusoid1}
v(t) &=V_\text{rms}\sqrt2\cos(\omega_0t+\varphi) \\
\label{eqn:mix:cos-sin-sinusoid2}
    &= V'\sqrt2\cos(\omega_0t) - V''\sqrt2\sin(\omega_0t)~,
\end{align}
with
\begin{align}
V' &= V_\text{rms}\cos\varphi\\
V'' &= V_\text{rms}\sin\varphi\\
V_\text{rms} &=\sqrt{(V')^2+(V'')^2}\\
\varphi&=\arctan(V''/V')~.
\end{align} 
The form \req{eqn:mix:cos-sin-sinusoid1}-\req{eqn:mix:cos-sin-sinusoid2}
relates to the \emph{phasor} representation\footnote{This is also known as the \emph{complex} representation, or as the \emph{Fresnel vector} representation.}\begin{align}
\label{eqn:mix:phasor-sinusoid}
V=V'+jV''=|V|e^{j\varphi}~,
\end{align}
which is obtained by freezing the $\omega_0$ oscillation, and by turning the amplitude into a complex quantity of modulus 
\begin{align}
|V|=\sqrt{(V')^2+(V'')^2} = V_\text{rms}
\end{align}
equal to the rms value of the time-domain sinusoid, and of argument 
\begin{align}
\varphi=\arctan\frac{V''}{V'}
\end{align}
equal to the phase $\varphi$ of the time-domain sinusoid.  The ``$\sin\omega_0t$'' term in Eq.~\req{eqn:mix:cos-sin-sinusoid2} has a sign ``$-$'' for consistency with Eq.~\req{eqn:mix:phasor-sinusoid}.

Another form frequently used is the \emph{analytic} (complex) signal  
\begin{align}
\label{eqn:mix:analytic-signal}
v(t) = Ve^{j\omega_0t}~,
\end{align}
where the complex voltage $V=V'+jV''$ is consistent with Eq.~\req{eqn:mix:phasor-sinusoid}.
The analytic signal has zero energy at negative frequencies, and double energy at positive frequencies.

\begin{figure}[t]
  \centering%     % COMBINED LATEX-PS/PDF FIGURES
	\centering\scalebox{\normalscale}{%  %              % scale
	\input{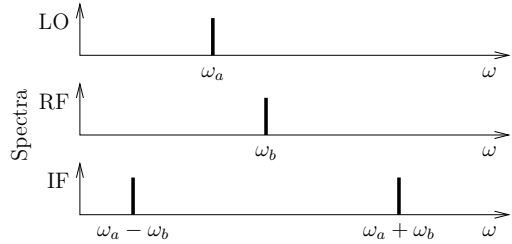}}
\caption{Frequency conversion.  Negative frequencies are not shown.}
\label{fig:mix-up-down-conversion}
\end{figure}

The product of two signals can only be described in the time domain [Eq.~\req{eqn:mix:simple-sinusoid}, \req{eqn:mix:cos-sin-sinusoid1}, \req{eqn:mix:cos-sin-sinusoid2}].
In fact, the phasor representation \req{eqn:mix:phasor-sinusoid} is useless, and the analytic signal \req{eqn:mix:analytic-signal} hides the down-conversion mechanism.   This occurs because 
$e^{j\omega_at}e^{j\omega_bt}=e^{j(\omega_a+\omega_b)t}$, 
while the product of two sinusoids is governed by 
\begin{align}
\label{eqn:mix:product-1}
\cos(\omega_at) \cos(\omega_bt) 
&= \frac12\cos\bigl(\omega_a-\omega_b\bigr)t+\frac12\cos\bigl(\omega_a+\omega_b\bigr)t\\
\label{eqn:mix:product-2}
\sin(\omega_at)  \cos(\omega_bt) 
&= \frac12\sin\bigl(\omega_a-\omega_b\bigr)t+\frac12\sin\bigl(\omega_a+\omega_b\bigr)t\\
\label{eqn:mix:product-3}
\sin(\omega_at)  \sin(\omega_bt)  
&= \frac12\cos\bigl(\omega_a-\omega_b\bigr)t-\frac12\cos\bigl(\omega_a+\omega_b\bigr)t~.
\end{align}
Thus, the product of two sinusoids yields the sum and the difference of the two input frequencies (Fig.~\ref{fig:mix-up-down-conversion}).
A pure sinusoidal signal is represented as a pair of Dirac delta function $\delta(\omega-\omega_0)$ and $\delta(\omega+\omega_0)$ in the spectrum, or as a single $\delta(\omega-\omega_0)$ in the case of the analytic signal.  
All the forms \req{eqn:mix:simple-sinusoid},  \req{eqn:mix:cos-sin-sinusoid1}, \req{eqn:mix:cos-sin-sinusoid2}, \req{eqn:mix:phasor-sinusoid}, and \req{eqn:mix:analytic-signal} are also suitable to represent (slow-varying) modulated signals. 
A modulated signal can be represented\footnote{The factor $\sqrt2$ is dropped, for $A$ is a peak amplitude.  Thus, $A'(t)$ and $A''(t)$ are the time-varying counterpart of $V'\sqrt2$ and $V''\sqrt2$.} as 
\begin{align}
\label{eqn:mix:modulated-sinusoid}
v(t)=A'(t)\cos(\omega_0t)-A''(t)\sin(\omega_0t)~.
\end{align}
$A'(t)$ and $A''(t)$ are the low-pass signals that contain information.  They may include a dc term, which accounts for the carrier, like in the old AM and PM\@.  Strictly, it is not necessary that $A'(t)$ and $A''(t)$ are narrow-band.
The time-depencence of $A'(t)$ and $A''(t)$ spreads the power around $\omega_0$.  The spectrum of the modulated signal is a copy of the two-side spectrum of $A'(t)$ and $A''(t)$ translated to $\pm\omega_0$. Thus, the bandwidth of the modulated signal \req{eqn:mix:modulated-sinusoid} is twice the bandwidth of $A'(t)$ and $A''(t)$.
Not knowing the real shape, the spectrum can be conventionally represented as a rectangle centered at the carrier frequency, which occupies the bandwidth of $A'$ and $A''$ on each side of $\pm\omega_0$ (Fig.~\ref{fig:mix-lc-conv}).  

Of course, Equations \req{eqn:mix:product-1}--\req{eqn:mix:product-3} also apply to the product of modulated signals, with their time-dependent coefficients $A'(t)$ and $A''(t)$.
Using mixers, we often encounter the product of a pure sinusoid [Eq.~\req{eqn:mix:simple-sinusoid}] multiplied by a modulated signal [Eq.~\req{eqn:mix:modulated-sinusoid}].  The spectrum of such product consists of two replicas of the modulated input, translated to the frequency sum and to the frequency difference (IF signal Fig.~\ref{fig:mix-lc-conv}).

%=============================================
\section{Linear modes}\label{sec:mix:lin-modes}
%=============================================
For the mixer to operate in any of the linear modes, it is necessary that
\begin{itemize}
\item the LO port is saturated by a suitable sinusoidal signal,
\item a small (narrowband) signal is present at the RF input.
\end{itemize}
The reader should refer to Sec.~\ref{ssec:mix:linearity} for more details about linearity.

%-------------------------------------------------------------------------------
\subsection{Linear frequency converter (LC) mode}\label{ssec:mix:lc-mode}
\begin{figure}[ht]
  \centering%     % COMBINED LATEX-PS/PDF FIGURES
	\centering\scalebox{\normalscale}{%  %              % scale
	\input{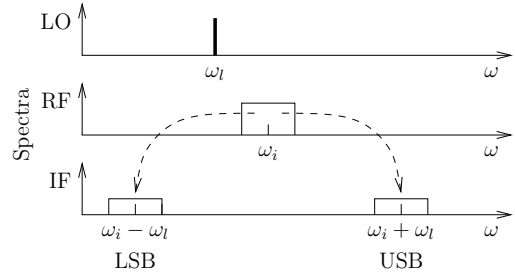}}
\caption{Frequency domain representation of the linear converter mode. Negative frequencies are not shown.}
\label{fig:mix-lc-conv}
\end{figure}

%------------------------------------------------------------------------------
The additional condition for the mixer to operate as a linear frequency converter is that the LO and the RF signals are separated in the frequency domain (Fig.~\ref{fig:mix-lc-conv}).

\begin{figure}[ht]
\centering%     % COMBINED LATEX-PS/PDF FIGURES
	\centering\scalebox{0.7}{%  %              % scale
	\input{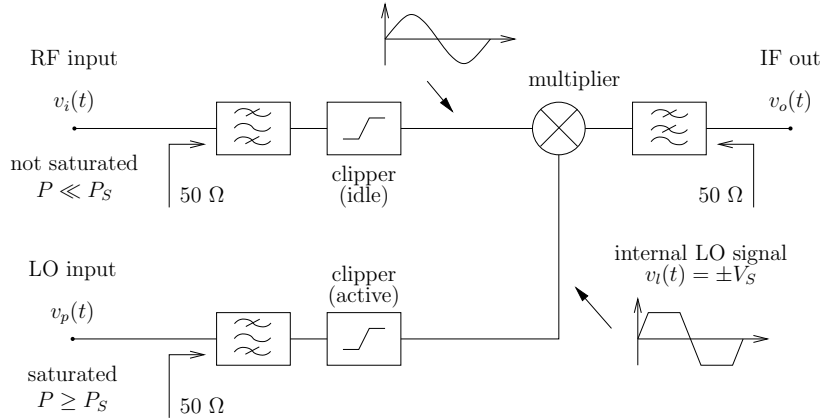}}
\caption{Model of the double balanced mixer operated
  as a linear converter.}
\label{fig:mix-lc-model}
\end{figure}

It is often convenient to describe the mixer as a system (Fig.~\ref{fig:mix-lc-model}), in which the behavior is modeled with functional blocks.
The clipper at the LO input limits the signal to the saturation level
$V_S$, while the clipper at the RF port is idle because this port is
not saturated.  The overall effect is that the internal LO voltage
$v_l(t)$ is approximately a trapezoidal waveform that switches between the
saturated levels ${\pm}V_S$.  The value of $V_S$ is a characteristic parameter of the specific mixer.  
The effect of higher LO power is to shrink the fraction of period taken by the slanted edges, rather than increasing $V_S$.
The asymptotic expression of $v_l(t)$ for strong saturation is
\begin{align}
v_l(t)&=\frac{4}{\pi}V_S\sum_{\mathrm{odd}~k\ge1}%
       \Big(\!-1\Big)^{\textstyle\frac{k-1}{2}}~~%
       \frac{1}{k}\;\cos(k\omega_lt)
\label{eqn:mix:multih-lo}\\
&=\frac{4}{\pi}V_S\left[\cos\omega_lt-\frac{1}{3}\cos3\omega_lt
	+\frac{1}{5}\cos5\omega_lt-\ldots+\ldots\;\right]
	\nonumber
\end{align}
The filters account for the bandwidth limitations of the actual mixer.
The IF output is often coupled in dc.  As an example,
Table~\ref{tab:mix-example} gives the main characteristics of two
typical mixers.
\begin{table}[ht]
\centering
\caption{Example of double balanced mixers.\dr}
\label{tab:mix-example}
\begin{tabular}{|c|c|c|}\hline
port & HF-UHF mixer         & microwave mixer     \\ \hline
LO   & 1--500 MHz           & 8.4--18 GHz  \\
     & 7 dBm $\pm1$ dB      & 8--11 dBm    \\
     & $\textsc{swr}<1.8$   & $\textsc{swr}<2$ \\\hline
RF   & 1--500 MHz           & 8.4--18 GHz  \\
     & 0 dBm max            & 0 dBm max    \\
     & $\textsc{swr}<1.5$   & $\textsc{swr}<2$ \\\hline
IF   & dc -- 500 MHz        & dc -- 2 GHz  \\
     & 0 dBm max            & 0 dBm max    \\
     & $\textsc{swr}<1.5$   & $\textsc{swr}<2$ \\
     & \textsc{ssb} loss 5.5 dB max & \textsc{ssb} loss 7.5 dB max \\\hline
\multicolumn{3}{|c|}{all ports terminated to 50 \ohm} \\\hline
\end{tabular}
\end{table}

A simplified description of the mixer is obtained by approximating the internal LO waveform $v_l(t)$ with the first term of its Fourier expansion
\begin{equation}
  v_l(t) = V_L\cos(\omega_lt)~~.
  \label{eqn:mix:lo} 
\end{equation}
The input signal takes the form
\begin{equation}
  v_i(t) = A_i(t)\cos\left[\omega_it+\varphi_i(t)\right]~~,
\end{equation}
where $A_i(t)$ and $\varphi_i(t)$ are the slow-varying signals in which
information is coded.  They may contain a dc term.  The output signal is
\begin{align}
v_o(t) & = \frac{1}{U}~v_i(t)\,v_l(t) \\
       & = \frac{1}{U}~A_i(t)\cos\bigl[\omega_it+\varphi_i(t)\bigr]%
           ~~V_L\cos(\omega_lt) \\
       & = \frac{1}{2U}\,V_LA_i(t)\:\Bigl\{
           \cos\bigl[(\omega_l-\omega_i)t-\varphi_i(t)\bigr]+
           \cos\bigl[(\omega_l+\omega_i)t+\varphi_i(t)\bigr]\Bigr\}~~.
\label{eqn:mix:lc-vo}
\end{align}
The trivial term $U=1$~V is introduced for the result to
have the physical dimension of voltage.

An optional bandpass filter, not shown in Fig.~\ref{fig:mix-lc-model},
may select the upper sideband (USB) or the lower sideband (LSB).  If
it is present, the output signal is
\begin{align}
v_o(t) &= \frac{1}{2U}\,V_LA_i(t)\,
          \cos\bigl[(\omega_l-\omega_i)t-\varphi_i(t)\bigr]
          \qquad\mbox{LSB} \\
v_o(t) &= \frac{1}{2U}\,V_LA_i(t)\,
          \cos\bigl[(\omega_l+\omega_i)t+\varphi_i(t)\bigr]
          \qquad\mbox{USB}~~.
\end{align}

\paragraph{Image frequency.}
%----------------------------------------------------------------
\begin{figure}[t]
\centering%     % COMBINED LATEX-PS/PDF FIGURES
	\centering\scalebox{\normalscale}{%  %              % scale
	\input{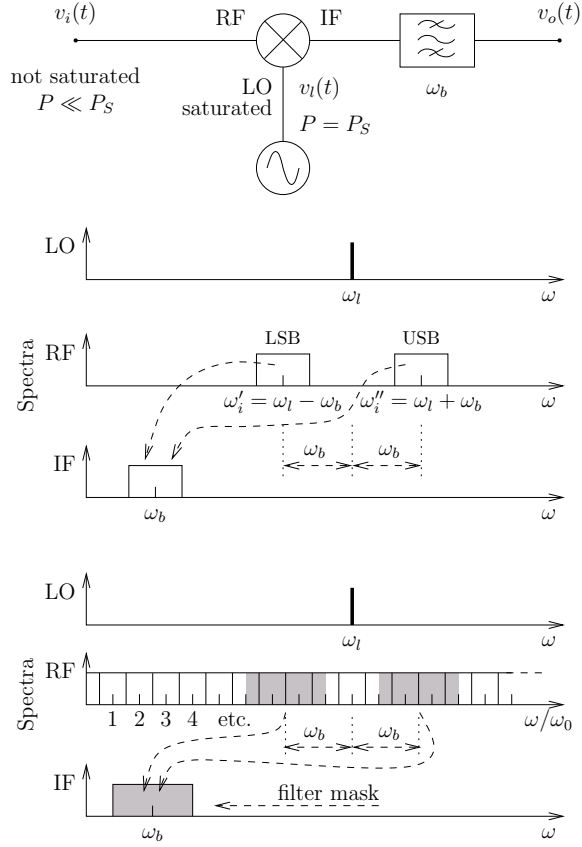}}
\caption{Image frequency in a conversion circuit.}
\label{fig:mix-lc-image}
\end{figure}
Let us now consider the inverse problem, that is, the identification of the input signal by observing the output of a mixer followed by a band-pass filter (Fig.~\ref{fig:mix-lc-image}~top).  In a typical case,  the output is a band-pass signal
\begin{equation}
v_o(t)  = A_o(t)\cos\big[\omega_bt+\varphi_o(t)\big]~~,
\label{eqn:mix:image}
\end{equation}
centered at $\omega_b$, close to the filter center frequency. 
It is easily proved that there exist two input signals
\begin{align}
v_L(t)  &= A_L(t)\cos\bigl[(\omega_l-\omega_b)t+\varphi_L(t)\bigr]&&\text{LSB}
\label{eqn:mix:image-1}\\
v_U(t) &= A_U(t)\cos\bigl[(\omega_l+\omega_b)t+\varphi_U(t)\bigr]&&\text{USB}~~,
\label{eqn:mix:image-2}
\end{align}
that produce a signal that passes through the output filter, thus
contribute to $v_o(t)$.  It is therefore impossible to ascribe a given
$v_o(t)$ to $v_L(t)$ or to its \emph{image} $v_U(t)$ if no a-priori
information is given.  Fig.~\ref{fig:mix-lc-image}~(middle) gives the explanation in terms of spectra.  
The USB and the LSB are image of one another with respect to $\omega_l$.  In most practical cases, one wants to detect one signal, so the presence of some energy around the image frequency is a nuisance.
In the case of the superheterodyne receiver, there results ambiguity in the frequency at which the receivers is tuned.  Even worse, a signal at the image frequency interferes with the desired signal. 
The obvious cure is a preselector filter preceding the mixer input.

More generally, the input signal can be written as 
\begin{equation}
v_i(t)  = \sum_n A_n'(t)\cos(n\omega_0t)-A_n''(t)\sin(n\omega_0t)~~,
\label{eqn:mix:image-3}
\end{equation}
which is a series of contiguous bandpass processes of bandwidth
$\omega_0$, centered around $n\omega_0$, and spaced by $\omega_0$.
The output is 
\begin{equation}
v_o(t)  = \frac{1}{U}\big[v_l(t)\,v_i(t)\big]*h_{bp}(t)~~,
\end{equation}
where ``$*$'' is the convolution operator, and $h_{bp}(t)$ the impulse
response of the bandpass IF filter.  The convolution  ${}*h_{bp}(t)$ defines the pass-band filtering.  Accordingly, the terms of $v_i(t)$ for which $|n\omega_0-\omega_l|$ is in the pass-band of the filter
contribute to the output signal $v_o(t)$.  Fig.~\ref{fig:mix-lc-image}~(bottom) shows the complete conversion process.

\begin{figure}[t]
\centering%     % COMBINED LATEX-PS/PDF FIGURES
	\centering\scalebox{\normalscale}{%  %              % scale
	\input{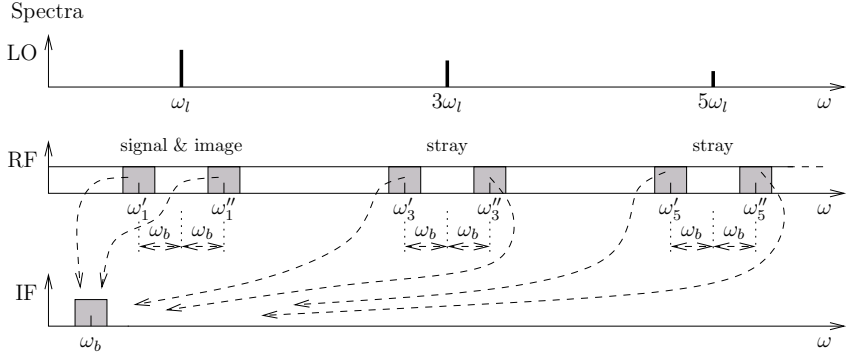}}
\caption{Multi-harmonic conversion.}
\label{fig:mix-lc-harm}
\end{figure}

\paragraph{Multi-harmonic conversion.}
%----------------------------------------------------------------
In usual conditions, the LO port is well saturated.  Hence it makes sense to account for several terms of the Fourier expansion \req{eqn:mix:multih-lo}
of the LO signal.
Each term of Eq.\ \req{eqn:mix:multih-lo} is a sinusoid of frequency
$k\omega_l$ that converts the portions of spectrum centered at 
$|k\omega_l+\omega_b|$ and $|k\omega_l-\omega_b|$ into $\omega_b$
(Fig.~\ref{fig:mix-lc-harm}), thus
\begin{align}
v_o(t) & = \frac{1}{U}~v_i(t)\,v_l(t) \\[1ex]
       & = \frac{1}{U}~A_i(t)\cos\bigl[\omega_it+\varphi_i(t)\bigr]%
        ~~\frac{4}{\pi}V_S\!\!\!\sum_{\mathrm{odd}~k\ge1}%
       \!\!\Big(-1\Big)^{\textstyle\frac{k-1}{2}}%
       \;\frac{1}{k}\,\cos(k\omega_lt) \\[1ex]
       & = \frac{1}{2U}\,\frac{4}{\pi}V_S\,A_i(t)%
           \!\!\!\sum_{\mathrm{odd}~k\ge1}%
           \!\!\!\Big(-1\Big)^{\textstyle\frac{k-1}{2}}%
	  \;\frac{1}{k}\Bigl\{
           \cos\bigl[(k\omega_l-\omega_i)t-\varphi_i(t)\bigr]+{}
           \Bigr.\nonumber\\
        &\hspace{32ex}\Bigl.
        {}+\cos\bigl[(k\omega_l+\omega_i)t+\varphi_i(t)\bigr]\Bigr\}~~.
\label{eqn:mix:lc-vo-multih}
\end{align}
With $k=1$, one term can be regarded as the signal to be detected, and the other one as the image.  All the terms with $k>1$, thus $3\omega_0$, $5\omega_0$, etc., as stray signals taken in because of distortion.  Of course, the mixer can be intentionally used to convert some frequency slot through multiplication by one harmonic of the LO, at the cost of lower conversion efficiency.  A bandpass filter at the RF input is often necessary to stop unwanted signals.  Sampling mixers are designed for this specific operation.  Yet their internal structure differs from that of the common double-balanced mixer.  

In real mixers the Fourier series expansion of $v_l(t)$ can be written as
\begin{equation}
v_l(t)=\sum_{\mathrm{odd}~k\ge1}%
       \Big(-1\Big)^{\textstyle\frac{k-1}{2}}%
       V_{L,k}\cos(k\omega_lt+\varphi_k)~~,
\label{eqn:mix:multih-lo-1}
\end{equation}
for Eq.~\req{eqn:mix:lc-vo-multih} becomes 
\begin{align}
v_o(t) & = \frac{1}{2U}\,A_i(t)%
           \!\!\!\sum_{\mathrm{odd}~k\ge1}%
           \!\!\!\Big(-1\Big)^{\textstyle\frac{k-1}{2}}%
	  V_{L,k}\,\Bigl\{
           \cos\bigl[(k\omega_l-\omega_i)t-\varphi_i(t)\bigr]+{}
           \Bigr.\nonumber\\
        &\hspace{32ex}\Bigl.
        {}+\cos\bigl[(k\omega_l+\omega_i)t+\varphi_i(t)\bigr]\Bigr\}~~.
\label{eqn:mix:lc-vo-multihreal}
\end{align}
The first term of Eq.~\req{eqn:mix:multih-lo-1} is equivalent to \req{eqn:mix:lo}, thus $V_{L,1}=V_L$.
Equation \req{eqn:mix:multih-lo-1} differs from Eq.~\req{eqn:mix:multih-lo} in the presence of the phase terms $\varphi_k$, and in 
that the coefficient $V_{L,k}$ decrease more rapidely than $1/k$.  This due to non-perfect saturation and to bandwidth limitation.  In weak saturation conditions the coefficient $V_{L,k}$ decrease even faster.

Looking at Eq.\ \req{eqn:mix:multih-lo}, one should recall that frequency multiplication results in phase noise multiplication.  If the LO signal contains a (random) phase $\phi(t)$, the phase $k\phi(t)$ is present in the $k$-th term.

For a more accurate analysis, the diode can no longer be modeled as a switch.
The diode forward current $i_F$ is governed by the exponential law
\begin{equation}
i_F=I_s\left(e^{\textstyle\frac{v_F}{\eta V_T}}-1\right)
\end{equation}
where $V_F$ is the forward voltage, $I_s$ the inverse saturation
current, $\eta\in[1\ldots2]$ a technical parameter of the junction, and
$V_T=kT/q$ the thermal voltage at the junction temperature.
At room temperature, it holds that $V_T=kT/q\simeq25.6$~mV\@.
The term ``$-1$'' is negligible in our case.  In the presence of a sinusoidal pump signal, the exponential diode current can be expanded using the identity
\begin{equation}
e^{z\cos\varphi} = I_0(z) 
                  +2\sum_{k=1}^{\infty}I_k(z)\cos(k\varphi)~~,
\end{equation}
where $I_k(\cdot)$ is the modified Bessel function of order $k$.  
As a consequence of the mixer symmetry, the even harmonics are canceled and the odd harmonics reinforced.  Ogawa~\cite{ogawa80mtt} gives an expression of the IF output current
\begin{equation}
i_o(t)=4I_s\frac{V\p{rf}}{\eta V_T}
       \sum_{\text{odd}~k\ge1} 
       I_k\left(\frac{V\p{lo}}{\eta V_T}\right)
       \Big[\cos(k\omega_l+\omega_i)t+\cos(k\omega_l-\omega_i)t\Big]~~.
\label{eqn:mix:ogawa}
\end{equation}
Equation \req{eqn:mix:ogawa} is valuable for design purposes.  Yet, it is of limited usefulness in analysis because some
parameters, like $I_s$ and $\eta$ are hardly available.
In addition, Eq.\ \req{eqn:mix:ogawa} holds in quasistatic conditions and does not account for a number of known effects, like stray inductances and capacitances, varactor effect in diodes, bulk resistance of the semiconductors, and other losses.  Nonetheless, Eq.\ \req{eqn:mix:ogawa} provides insight in the nature of the coefficients $V_{L,k}$.

\paragraph{Rules for the load impedance at the IF port.}\label{par:mix-if-load-impedance}
%----------------------------------------------------------------
The product of two sinusoids at frequency $\omega_i$ and $\omega_l$, inherently, contains the frequencies $\omega_i\pm\omega_l$. 
At the IF port, current flow must be allowed at both these frequencies, otherwise the diodes can not switch. The problem arises when IF selection filter shows high impedance in the stop band.  Conversely, low impedance $Z\ll R_0$ is usually allowed.
\begin{figure}[t]
\centering%     % COMBINED LATEX-PS/PDF FIGURES
	\centering\scalebox{\normalscale}{%  %              % scale
	\input{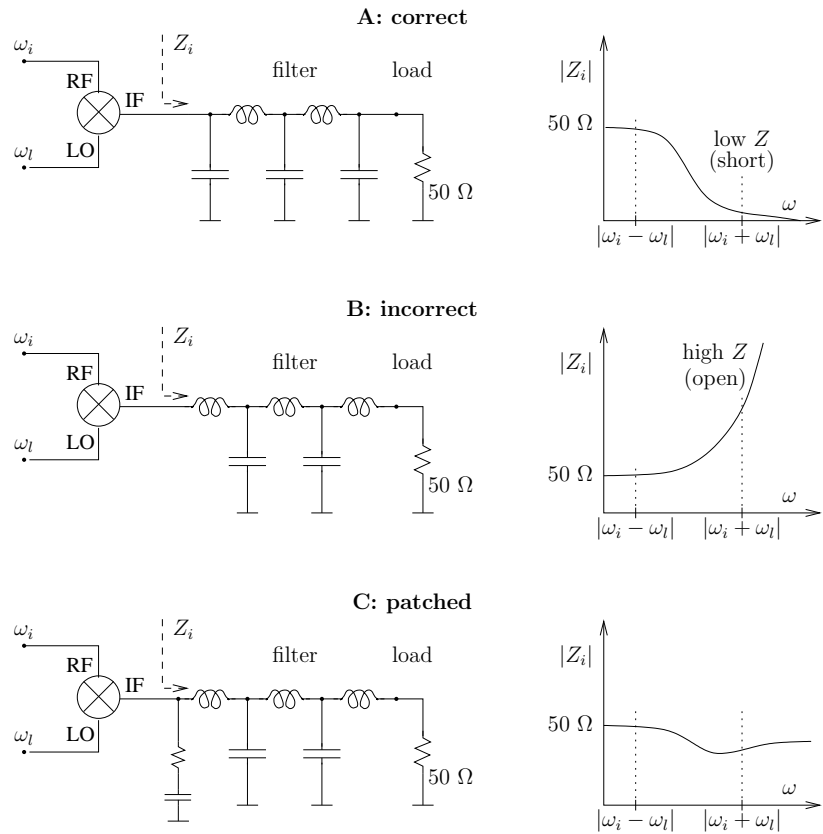}}
\caption{The mixer is followed by a filter that selects the $|\omega_i-\omega_l|$ frequency.}
\label{fig:mix-if-filter}
\end{figure}
Figure \ref{fig:mix-if-filter} shows three typical cases in which a filter is used to select the $|\omega_i-\omega_l|$ signal at the IF output, and to reject the image at the frequency $|\omega_i+\omega_l|$.
The scheme A is correct because the image-frequency current can flow through the diodes (low impedance).  The scheme B will not work because the filter is nearly open circuit at the image frequency.  The scheme C is a patched version of B, in which an additional $RC$ cell provides the current path for the image frequency.
The efficient use of a mixer as a multi-harmonic converter may require a specific analysis of the filter.  

In microwave mixers, the problem of providing a current path to the image frequency may not be visible, having been fixed inside the mixer.
This may be necessary when the image frequency is out of the bandwidth, for the external load can not provide the appropriate impedance.

Rules are different in the case of the \emph{phase detector} because the current path is necessary at the $2\omega_l$ frequency, not at dc.

\paragraph{Can the LO and RF ports be interchanged?}
%----------------------------------------------------------------
With an ideal mixer yes, in practice often better not.  
Looking at Fig.~\ref{fig:mix-dbm}, the center point of the LO transformer is
grounded, which helps isolation.  In the design of microwave mixers,
where the transformers are replaced with microstrip baluns,
optimization may privilege isolation from the LO pump, and low loss in
the RF circuit.
This is implied in the general rule that the mixer is designed and documented for the superheterodyne receiver.   Nonetheless, interchanging RF and LO can be useful in some cases, for example to take benefit from the difference in the input bandwidth.

%-------------------------------------------------------------------------------
\subsection{Linear Synchronous Detector (SD) Mode}\label{ssec:mix:sd-mode}
%-------------------------------------------------------------------------------
\begin{figure}[t]
  \centering%     % COMBINED LATEX-PS/PDF FIGURES
	\centering\scalebox{\normalscale}{%  %              % scale
	\input{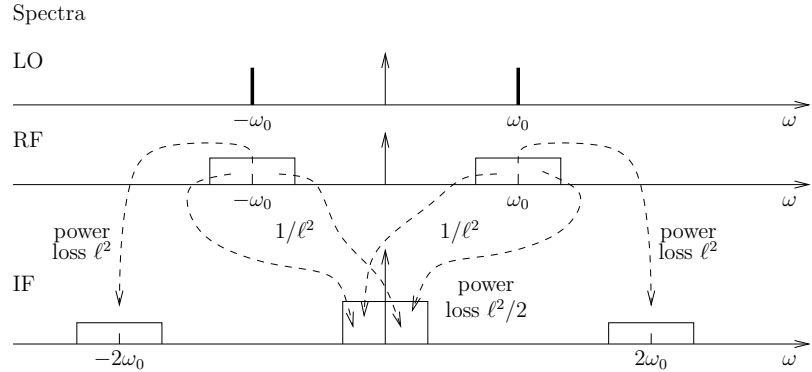}}
\caption{Frequency-domain sketch of the linear synchronous 
  detection.}
\label{fig:mix-sd-conv}
\end{figure}
The general conditions for the linear modes are that the LO port is saturated by a suitable sinusoidal signal, and that a small (narrowband) signal is present at the RF input.  The additional conditions for the mixer to operate in the SD mode  are: (1) the LO frequency $\omega_l$ is tuned at the center of the spectrum of the (narrowband) RF signal, and (2) the IF output is low-passed.

The basic mixer operation is the same of the frequency conversion
mode, with the diode ring used as a switch that inverts or not the
input polarity dependig on the sign of the LO\@.  The model of
Fig.~\ref{fig:mix-lc-model} is also suitable to the SD mode.  Yet, the
frequency conversion mechanism is slightly different.
Figure~\ref{fig:mix-sd-conv} shows the SD mode in the frequency
domain, making use of two-sided spectra.  Using one-sided spectra, the
conversion products of negative frequency are folded to positive
frequencies.
\begin{figure}[t]
\centering%     % COMBINED LATEX-PS/PDF FIGURES
	\centering\scalebox{\normalscale}{%  %              % scale
	\input{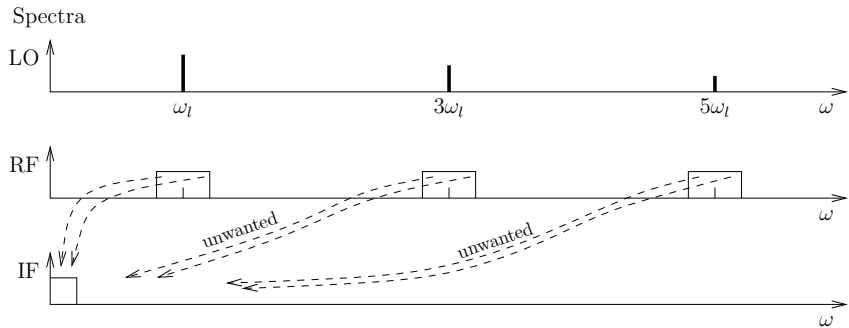}}
\caption{Signals are converted to IF by the harmonics at frequency multiple than the LO frequency.}
\label{fig:mix-ld-harm}
\end{figure}
Of course, the multi-harmonic frequency conversion mechanism, due to the harmonics multiple of the LO frequency still works (Figure~\ref{fig:mix-ld-harm}).

\begin{figure}[t]
  \centering%     % COMBINED LATEX-PS/PDF FIGURES
	\centering\scalebox{\normalscale}{%  %              % scale
	\input{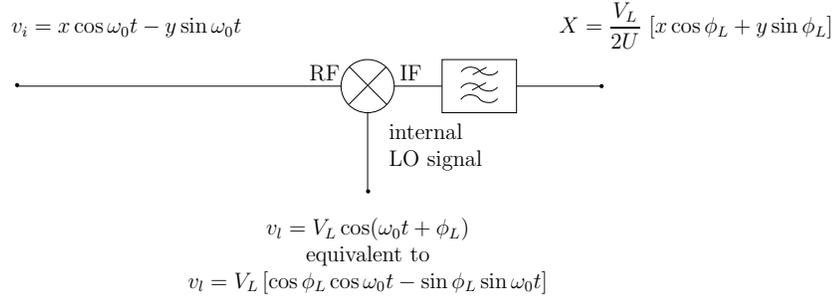}}
\caption{Linear synchronous detection.}
\label{fig:mix-lc-scalar}
\end{figure}
The simplest way to understand the synchronous conversion is to
represent the input and the internal LO signal $v'_l(t)=V_L\cos(\omega_0t+\varphi_L)$ in Cartesian coordinates\footnote{In this Section we use $x$ and $y$ in order to emphasize some properties of the synchronous detection tightly connected to Cartesian-coordinate representation.  Here, $x$ and $y$ are the same thing of $A'$ and $A''$ of Eq.~\req{eqn:mix:modulated-sinusoid}.}
\begin{align}
\label{eqn:mix:x-y-signal}
v_i(t)&=x(t)\cos\omega_0t-y(t)\sin\omega_0t\\
v'_l(t)&=V_L\left[\cos\varphi_L\cos\omega_0t-\sin\varphi_L\sin\omega_0t\right]
\end{align}
The signal at the output of the low-pass filter is\footnote{Once again, we emphasize the properties connected with the Cartesian-coordinate representation.  $X(t)$ is the same thing of $v_o(t)$ of other sections.} (Fig.~\ref{fig:mix-lc-scalar})
\begin{align}
X(t) 
& = \frac{1}{U}\,v_i(t)\,v'_l(t) * h_{lp}\\
& = \frac{1}{U}\,\bigl[x\cos\omega_0t-y\sin\omega_0t\bigr]
      \:V_L\bigl[\cos\varphi_L\cos\omega_0t-\sin\varphi_L\sin\omega_0t\bigr]*h_{lp}\\
& = \frac{1}{2U}\,V_L
      \Big[x\cos\varphi_L+y\sin\varphi_L + \text{($2\omega$ terms)}\Big]*h_{lp}~,\\
\intertext{thus,}      
X(t) 
&= \frac{1}{2U}\,V_L
           \Big[x(t)\cos\varphi_L+y(t)\sin\varphi_L\Big]~~.
\label{eqn:mix:scalar-product-x}
\end{align}
Eq.~\req{eqn:mix:scalar-product-x} can be interpreted as the scalar product 
\begin{align}
X = \frac{1}{2U}\,V_L (x,y)\cdot(\cos\varphi_L,\sin\varphi_L)~,
\end{align}
plus a trivial factor $\frac{1}{2U}V_L$ that accounts for losses.

Let us now replace the LO signal $v'_l(t)$ with 
\begin{align}
v''_l(t)=-V_L\sin(\omega_0t+\varphi_L) = 
         -V_L\left[\sin\varphi_L\cos\omega_0t-\cos\varphi_L\sin\omega_0t\right]~.
\end{align}
In this conditions, the output signal is 
\begin{align}
Y(t) 
& = \frac{1}{U}\,v_i(t)\,v''_l(t) * h_{lp}\\
& = \frac{1}{U}\,\bigl[x\cos\omega_0t-y\sin\omega_0t\bigr]
      \:V_L\bigl[-\sin\varphi_L\cos\omega_0t-\cos\varphi_L\sin\omega_0t\bigr]*h_{lp}\\
& = \frac{1}{2U}\,V_L
      \Big[-x\sin\varphi_L+y\cos\varphi_L + \text{($2\omega$ terms)}\Big]*h_{lp}~,\\
\intertext{thus,}      
Y(t) &= \frac{1}{2U}\,V_L\Big[-x(t)\sin\varphi_L+y(t)\cos\varphi_L\Big]~~.
\label{eqn:mix:scalar-product-y}
\end{align}%
\begin{figure}[t]
\centering%     % COMBINED LATEX-PS/PDF FIGURES
	\centering\scalebox{\normalscale}{%  %              % scale
	\input{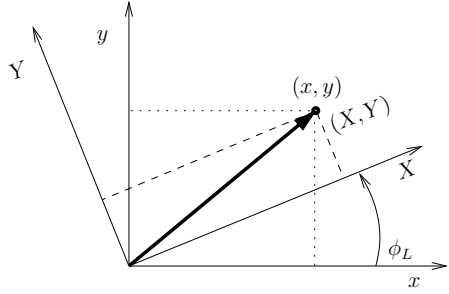}}
\caption{Cartesian-frame rotation.  The coefficient $\frac{1}{2U}V_L$ is implied.}
\label{fig:mix-frame-rotation}
\end{figure}%
Finally, by joining Equations \req{eqn:mix:scalar-product-x} and \req{eqn:mix:scalar-product-y}, we find
\begin{align}
\label{eqn:mix:iq-frame-rotation}
\left[\begin{array}{c}X(t)\\Y(t)\end{array}\right] & = 
\frac{1}{2U}\,V_L
\left[\begin{array}{cc}\cos\varphi_L&\sin\varphi_L\\-\sin\varphi_L&\cos\varphi_L\end{array}\right] 
\left[\begin{array}{c}x(t)\\y(t)\end{array}\right]~.
\end{align}
Equation~\req{eqn:mix:iq-frame-rotation} is the common form of a frame rotation by the angle $\varphi_L$ in Cartesian coordinates (Fig.~\ref{fig:mix-frame-rotation}).  

\begin{figure}[t]
  \centering%     % COMBINED LATEX-PS/PDF FIGURES
	\centering\scalebox{\normalscale}{%  %              % scale
	\input{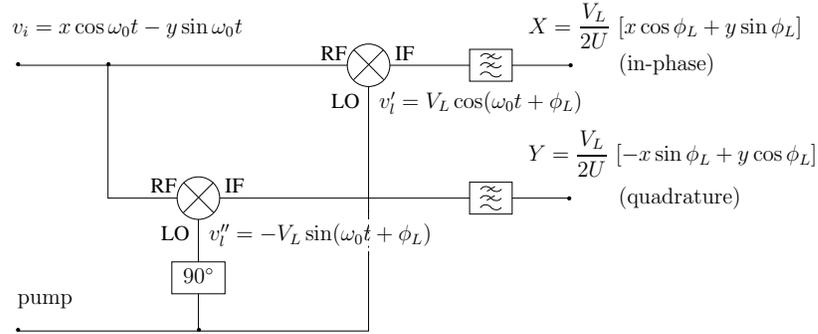}}
\caption{Basic I-Q detector.}
\label{fig:mix-lc-scalar-iq}
\end{figure}

The simultaneous detection of the input signal with two mixers pumped in quadrature is common in telecommunications, where QAM modulations are widely used\footnote{For example, the well known wireless standard 811g (WiFi) is a 64 QAM\@.  The transmitted signal is of the form \req{eqn:mix:x-y-signal}, with $x$ and $y$ quantized in 8 level (3 bits) each.}.
The theory of coherent communication is analyzed in \cite{viterbi:communication}.
Devices like that of Fig.~\ref{fig:mix-lc-scalar-iq}, known as I-Q detectors, are commercially available from numerous manufacturers.  Section~\ref{sec:mix:specials-iqs} provide more details on these devices.

%------------------------------------------------------------------------------
\subsection{Linearity}\label{ssec:mix:linearity}
%------------------------------------------------------------------------------
A function $f(\cdot)$ is said linear \cite{rudin:mathematical-analysis} if it has the following two properties
\begin{gather}
f(ax)=af(x)\\
f(x+y)=f(x)+f(y)~.
\end{gather}
The same definition applies to operators.
When a sinusoidal signal of appropriate power and frequency is sent to the LO port, \emph{the mixer is linear}, that is, \emph{the output signal $v_o(t)$ is a linear function of the input $v_i(t)$}.  This can be easily proved for the case of simple conversion [Eq.~\req{eqn:mix:lc-vo}]
\begin{align}
v_o(t) & = \frac{1}{2U}\,V_LA_i(t)\:\Bigl\{
           \cos\bigl[(\omega_l-\omega_i)t-\varphi_i(t)\bigr]+
           \cos\bigl[(\omega_l+\omega_i)t+\varphi_i(t)\bigr]\Bigr\}
           \nonumber
\end{align}
The linearity of $v_o(t)$ vs.\ $v_i(t)$ can also be demonstrated in the case of the multi-harmonic conversion, either by taking a square wave as the LO internal signal [Eq.~\req{eqn:mix:lc-vo-multih}], or by using the internal LO signal of real mixers [Eq.~\req{eqn:mix:lc-vo-multihreal}].  In fact, the Fourier series is a linear superposition of sinusoids, each of which treated as above.
In practice, the double balanced mixer can be used in a wide range of
frequency (up to $10^4$), where it is linear in a wide range of power, which may exceed $10^{16}$ (160 dB).

In large-signal conditions, the mixer output signal can be expanded as
the polynomial
\begin{equation}
v_o(v_i) = a_0 + a_1v_i + a_2v_i^2 + a_3v_i^3 + \ldots~~.
\label{eqn:mix-nonlinear-polynomial}
\end{equation}
The symmetric topology cancels the even powers of $v_i$, for the above
polynomial can not be truncated at the second order.  Yet, the
coefficient $a_2$ is nonzero because of the residual asymmetry in the
diodes and in the baluns.  Another reason to keep the third-order term
is the adjacent channel interference.  In
principle, transformer nonlinearity should also be analyzed.  In
practice, this problem is absent in microwave mixers, and a minor
concern with ferrite cores.  The coefficient $a_1$ is the invese loss
$\ell$.  The coefficients $a_2$ and $a_3$ are never given explicitely.
Instead, the intercept power (IP2 and IP3) is given, that is, the
power at which the nonlinear term ($a_2v_i^2$ and $a_3v_i^3$) is equal
to the linear term.

%==============================================
\section{Mixer loss}\label{sec:mix:loss}
\begin{figure}[t]
\centering%     % COMBINED LATEX-PS/PDF FIGURES
	\centering\scalebox{\normalscale}{%  %              % scale
	\input{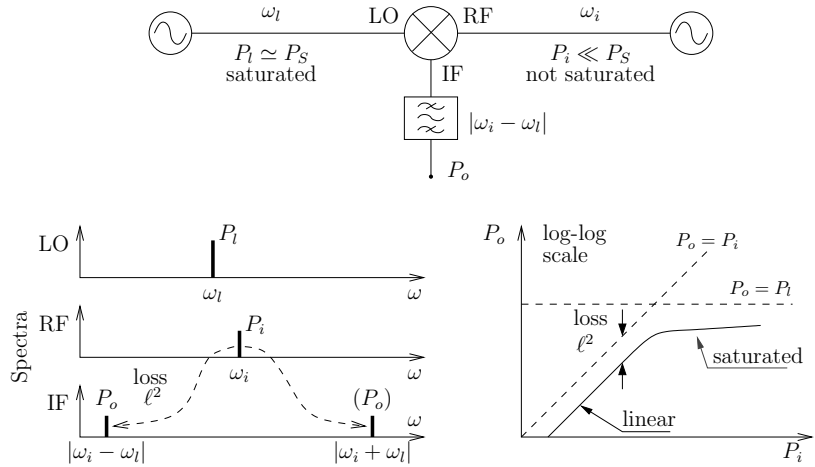}}
\caption{Definition of the SSB loss $\ell$.}
\label{fig:mix-loss}
\end{figure}

%==============================================
The conversion efficiency of the mixer is operationally defined via
the two-tone measurement shown in Fig.\ \ref{fig:mix-loss}.  
This is the case of a superheterodyne receiver in which the incoming signal is an unmodulated sinusoid $v_i(t)=V_i\cos\omega_it$, well below saturation.  The LO sinusoid is set to the nominal saturation power.  
In this condition, and neglecting the harmonic terms higher than the first, the output signal consists of a pair of sinusoids of frequency $\omega_o=|\omega_l\pm\omega_i|$.   One of these sinusoids, usually
$|\omega_l-\omega_i|$ is selected.  The SSB power loss $\ell^2$ of
the mixer is defined\footnote{In our previous articles we took $\ell=P_i/P_o$ instead of $\ell^2=P_i/P_o$.  The practical use is unchanged because $\ell$ is always given in dB.} as
\begin{equation}
\frac{1}{\ell^2}=\frac{P_o}{P_i}
\qquad\text{SSB loss $\ell$}
\label{eqn:mix-ssb-loss-def}
\end{equation}
where $P_i$ is the power of the RF input, and $P_o$ is the power of the IF output at the selected freqency.  The specifications of virtually all mixes resort to this definition.

The loss is about constant in a wide range of power and frequency.
The upper limit of the RF power range is the saturation power,
specified as the compression power $P_{1\unit{dB}}$ at which the loss
increases by 1 dB\@.

\paragraph{Intrinsic SSB loss.}
%----------------------------------------------------------------
The lowest loss refers to the ideal case of the zero-threshold diode, free from resistive dissipation. 
The LO power is entirely wasted in switching the diodes.  Under this assumptions, the ring of Figure~\ref{fig:mix-dbm} works as a loss-free switch that inverts or not the polarity of the RF, $v_o(t)=\pm v_i(t)$, according to the sign of $v_l(t)$.  Of course, the instantaneous output power is conserved
\begin{align}
\frac{1}{R_0}\:v_i^2(t) &= \frac{1}{R_0}\:v_o^2(t)~~. 
\end{align}
Nonetheless, the mixer splits the input power into the conversion products at frequency $|\omega_i\pm\omega_l|$ and higher harmonics, for only a fraction of the input power is converted into the desired frequency.
There result a loss \emph{inherent} in the frequency conversion process, found with the definition \req{eqn:mix-ssb-loss-def}.

In the described conditions, the internal LO signal is a unit square wave ($V_S=1$ V), whose Fourier series expansion is
\begin{equation}
v_l(t) = \frac{4}{\pi}\left[
  \cos\omega_lt-\frac{1}{3}\cos3\omega_lt
  +\frac{1}{5}\cos5\omega_lt -\ldots+\ldots\right]~~.
\end{equation}
Only the first term of the above contributes to the down-converted
signal at the frequency $\omega_b=|\omega_i-\omega_l|$.  
The peak amplitude of this term is $V_L=\frac{4}{\pi}$ V\@.  Hence,
\begin{align}
  v_o(t)
  &= \frac{1}{U}\,v_l(t)\,v_i(t) \\
  &= \frac{4}{\pi}\cos(\omega_lt) \; V_i\cos(\omega_it)\\
  &= \frac{4}{\pi} V_i \; \frac{1}{2}\Bigl\{
     \cos[(\omega_i-\omega_l)t]+\cos[(\omega_i+\omega_l)t]\Bigr\}\\
  &= \frac{2}{\pi}V_i \; \cos[\omega_bt] \qquad\mbox{rubbing out
    the USB}
\end{align}
The RF and IF power are 
\begin{equation}
P_i=\frac{V_i^2}{2R_0} \qquad\mbox{and}\qquad 
P_o=\frac{1}{2R_0}\,\frac{4V_i^2}{\pi^2} 
\end{equation}
from which the minimum loss $\ell=\sqrt{P_i/P_o}$ is
\begin{equation}
\ell=\frac{\pi}{2}~~\simeq1.57~~\text{(3.92 dB)}
\qquad\text{minimum SSB loss}.
\end{equation}

\begin{figure}[t]
\centering%     % COMBINED LATEX-PS/PDF FIGURES
	\centering\scalebox{\normalscale}{%  %              % scale
	\input{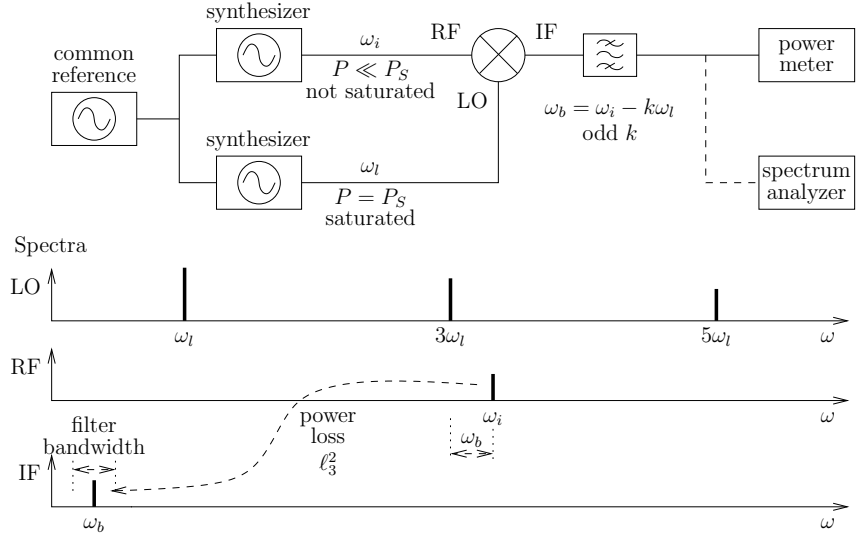}}
\caption{Measurement of the mixer loss in harmonic coversion.}
\label{fig:mix-loss-harm}
\end{figure}

\paragraph{SSB loss of actual mixers.}
%----------------------------------------------------------------
The loss of microwave mixer is usually between 6 dB for the 1-octave
devices, and 9 dB for 3-octave units.  The difference is due to the
microstrip baluns that match the nonlinear impedance of the diodes to
the 50 \ohm\ input over the device bandwidth.  In the case of a narrow-band mixer optimized for conversion efficiency, the SSB loss can be of 4.5 dB \cite{ogawa80mtt}.  The loss of most HF/UHF mixers is of about 5--6 dB in a band up to three decades.  This is due to the low loss and to the large bandwidth of the tranmission-line transformers.  Generally,
the LO saturation power is between 5 and 10 mW (7--10 dBm).  Some
mixers, optimized for best linearity make use of two or three diodes
in series, or of two diode rings (see Fig.\ \ref{fig:mix-multidiode}), 
and need larger LO power (up to 1 W).  The advantage of these mixers is high intercept power, at the cost of larger loss (2--3 dB more).
When the frequencies multiple of the LO frequency are exploited to convert the input signal, it may be necessary to measure the conversion loss.  A scheme is proposed in Fig.\ \ref{fig:mix-loss-harm}.

\paragraph{Derivation of the internal LO voltage from the loss.}
%----------------------------------------------------------------
For the purpose of analytical calculus, the amplitude $V_L$ of the internal LO signal is often needed.   With real (lossy) mixers, it holds that $V_L<\frac{4}{\pi}$ V\@.  $V_L$ can be derived by equating the
output power $P_i/\ell^2$ to the power of the output product.  
The usefulness of this approach is in that $\ell$ is always specified.  Let
\begin{equation}
v_i(t)=V_i\cos\left[\omega_i(t)+\varphi_i\right]
\end{equation}
the RF input, and select the lower\footnote{Some experimental
  advantages arise from taking $\omega_b=|\omega_i-\omega_l|$ instead
  of $\omega_b=|\omega_i+\omega_l|$.} output frequency $\omega_b=|\omega_i-\omega_l|$.
The internal LO signal is
\begin{equation}
v_l(t)=V_L\cos(\omega_lt+\varphi_l)~~.
\end{equation}
Measuring the output power, we can drop the phases $\varphi$ and $\varphi_l$.  Hence, the output signal is
\begin{align}
v_o(t)&=\frac{1}{U}V_iV_L
	\big[\cos\omega_it+\cos\omega_lt\big]*h_{bp}(t)\\
	&=\frac{1}{U}V_iV_L \frac{1}{2}\cos(\omega_i-\omega_l)t
\end{align}
The output power is
\begin{equation}
P_o=\frac{1}{2R_0} \; \frac{1}{4U^2}V_i^2V_L^2~~
\label{eqn:mix:po-product}
\end{equation}
when the input power is
\begin{equation}
P_i=\frac{1}{2R_0}\,V_i^2~~.
\end{equation}
Combining the two above Equations with the definition of
$\ell$ [Eq.\ \req{eqn:mix:po-product}], we obtain
\begin{equation}
\frac{1}{\ell^2} \: \frac{1}{2R_0}A_i^2  =
\frac{1}{2R_0} \; \frac{1}{4U^2}A_i^2V_L^2~~,
\end{equation}
hence 
\begin{equation}
V_L = \frac{2U}{\ell}\qquad\text{Internal LO peak amplitude}.
\label{eqn:mix:equiv-lo-v}
\end{equation}
Interestingly, the loss of most mixers is close to 6 dB, for
$V_L\simeq1$ V, while the intrinsic loss $\ell=\pi/2$ yields
$V_L=4/\pi\simeq1.27$~V\@.

\paragraph{What if the LO power differs from the nominal power?}
%----------------------------------------------------------------
\begin{figure}[t]
%\centering\includegraphics[bb=250 20 220 580, scale=0.6]{mix-p12}
\centering%     % COMBINED LATEX-PS/PDF FIGURES
	\centering\scalebox{\smallscale}{%  %              % scale
	\input{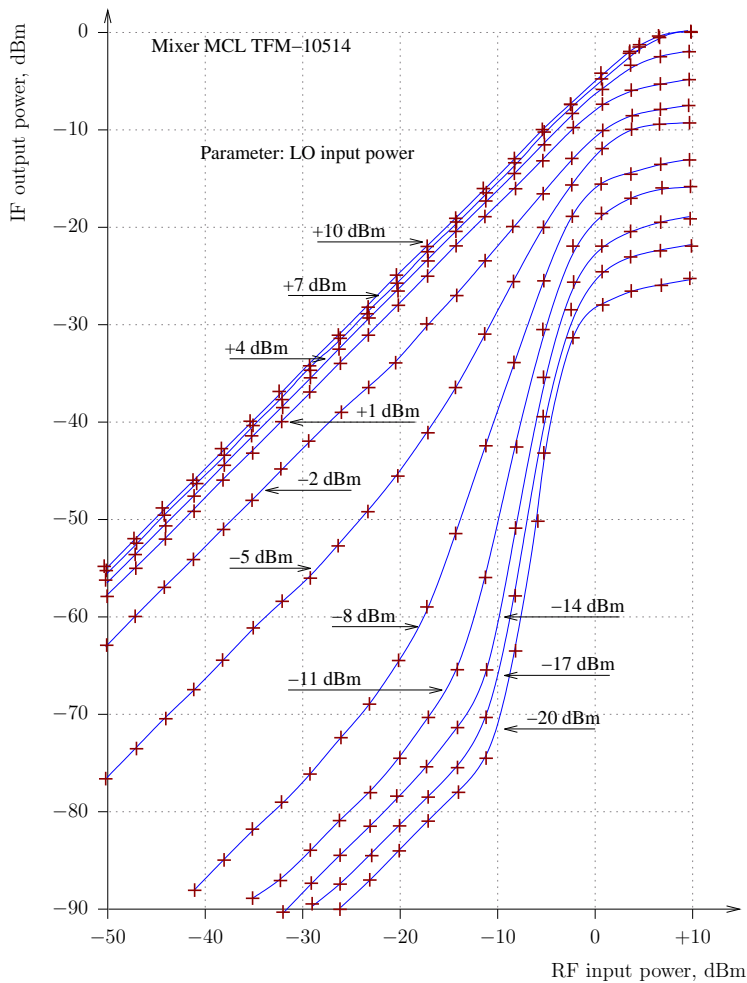}}
\caption{Conversion loss measured at various LO 
  power levels (1990 p.~12).}
\label{fig:mix-if-vs-lo-power}
\end{figure}
When the LO input is saturated, the LO power has little or no effect on the output signal.  
This fact is often referred as \emph{power desensitization} (also LO desensitization, or pump desensitization).  In a narrow power range, say ${\pm}2$ dB from the nominal power, the conversion loss changes slightly, and noise also varies.  The internal Schottky diodes exhibit exponential $i=i(v)$ characteristics, hence lower LO power is not sufficient to 
saturate the diodes, and the the ring is unable to switch.  The conversion
efficiency $1/\ell$ is reduced, and drops abruptly some 10 dB below the nominal LO power.  As a side effect of loss, white
noise increases.  Figure~\ref{fig:mix-if-vs-lo-power} shows an example of output
power as a function of the RF power, for various LO power levels.
Below the nominal LO power, flicker noise increases.  Whereas
this phenomenon is still unclear, we guess that this is due to the
increased fraction of period in which the diodes are neither open
circuit or saturated, and that up conversion of the near-dc flickering
of the junction takes place during this transition time .

Insufficient LO power may also impair symmetry, and in turn the
cancellation of even hamonics.  The physical explanation is that saturated
current is limited by the diode bulk resistance, which is more
reproducible than the exponential law of the forward current.
Increasing the fraction of time in which the exponential law dominates
emphasizes the asymmetry of the diodes.

Too high LO power may increase noise, and damage the mixer.  Special
care is recommanded with high-level mixers, in which the nominal LO
power of of 50 mW or more, and in the miniaturized mixers, where the small size limits the heat evacuation.

According to the model of Fig.~\ref{fig:mix-lc-model}, the LO clipper
limits the internal voltage to ${\pm}V_S$, which turns the input
sinusoid into a trapezoidal waveform.  Hence, the input power affects
the duration of the wavefronts, and in turns the harmonic contents.
As a result, a circuit may be sensitive to the LO power if stray input
signals are not filtered out properly.

Finally, changing the LO power affects the dc voltage at the IF
output.  This can be a serious problem when the mixer is used as a
synchronous converter or as a phase detector.

%================================================================
\section{Saturated Modes}\label{sec:mix:saturated-modes}
%================================================================
%
When both RF and LO inputs are saturated, the mixer behavior changes
radically.  The mixer can no longer be described as a simple switch
that invert or not the RF signal, depending on the LO sign.  Instead,
at each instant the largest signal controls the switch, and sets the
polarity of the other one.  Of course, the roles are interchanged
continuously.  Strong odd-order harmonics of the two input frequencies
are present, while even-order harmonics are attenuated or cancelled by
symmetry.  Saturation means that amplitude has little effect on the
output, for saturated modes are useful in phase detectors or in
frequency synthesis, where amplitudes are constant.  A further
consequence of saturation is phase noise multiplication, which is
inherent in harmonic generation.  In the case of saturated modes,
phase noise multiplication takes place in both LO and RF.

In saturated modes the specified maximum power at the RF port is
always exceeded.  When this maximum power is exceeded, the mixer
leaves the ``normal'' linear operation, still remaining in a safe
operating range until the ``absolute maximum ratings'' are approached.
Read page \pageref{sec:mix:safe-op}.

The model of Fig.~\ref{fig:mix-lc-model} describes some
characteristics, as it emphasizes the internally clipped waveforms,
and the cancellation of even harmonics.  Yet, the model fails in
predicting amplitude because the ring is no longer a multiplier.  The
output amplitude is lower than expected.

%================================================================
\subsection{Saturated Frequency Converter (SC) Mode}\label{sec:mix:sc-mode}
%================================================================
%
The conditions for the mixer to operate in SC mode are
\begin{itemize}\item the LO and the RF ports are saturated by sinusoidal signals,
\item the input frequencies are not equal, and the ratio
  $\omega_l/\omega_i$ is not too close to the the ratio of two small
  integers (say, 5--7),
\item the output is band-passed.
\end{itemize}%
Let the input signals 
\begin{align}
v_i(t) & = V'_P\cos\omega_it\\
v_p(t) & = V''_P\cos\omega_lt~~.
\end{align}
If possible, the saturated amplitudes $V'_P$ and $V''_P$ should be
equal.  The main output signal consists of the pair of sinusoids
\begin{equation}
v_o(t)=V_O\cos(\omega_l-\omega_i)t+V_O\cos(\omega_l+\omega_i)t
\label{eqn:mix:sfc-vo}
\end{equation}
that derives from the product $v_i(t)\,v_l(t)$.  Yet, the output
amplitude $V_O$ is chiefly due to the internal structure of the mixer,
and only partially influenced by $V'_P$ and $V''_P$.  A bandpass
filter selects the upper or the lower frequency of
\req{eqn:mix:sfc-vo}.

The unsuitability of the model of Fig.~\ref{fig:mix-lc-model} to
predict amplitude can be seen in the following example.
 
\begin{example}
  Replacing $V'_P$ and $V''_P$ with $V_L$ yields
  $V_O=\frac{1}{2}UV_L^2$.  Let us consider typical mixer that has a
  loss of 6 dB when the LO has the nominal power of 5 mW (7~dBm).
  From Eq.~\req{eqn:mix:equiv-lo-v} we get $V_L\simeq1$~V, thus we
  expect $V_O=250$ mV, and an output power $V_O^2/2R_0=2.5$ mW ($+4$
  dBm) with $R_0=50$~\ohm.  Yet, the actual power is hardly higher
  than 1.25~mW ($+1$~dBm).
\end{example}

\begin{figure}[t]
\centering%     % COMBINED LATEX-PS/PDF FIGURES
	\centering\scalebox{\normalscale}{%  %              % scale
	\input{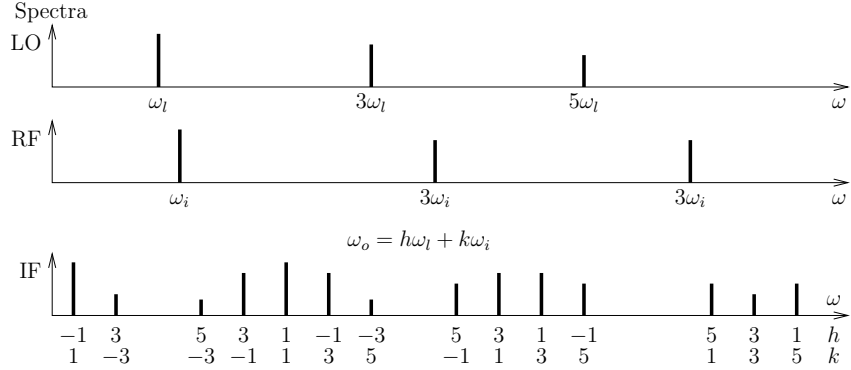}}
\caption{Frequency conversion with a saturated mixer.}
\label{fig:mix-sc-harm}
\end{figure}

Accounting for the harmoncs, the output signal is
\begin{equation}
v_o(t)=\sum_{\text{odd}~h,k}V_{hk}\cos(h\omega_l+k\omega_i)t
\qquad\begin{array}{c}
\text{\small positive frequencies}\\[-0.5ex]
\omega_{hk}=h\omega_l+k\omega_i>0
\end{array}~~,
\label{eqn:mix:sfc-vo-harm}
\end{equation}
where the sum is extended to the positive output frequencies, i.e.,
$h\omega_l+k\omega_i>0$.  $V_{hk}$ decreases more rapidely than the
product $|hk|$, and drops abruptly outside the bandwidth.
Figure~\ref{fig:mix-sc-harm} shows an example of spectra involving
harmonics.

The contition on the ratio $\omega_l/\omega_i$ two output frequencies
$\omega_{h'k'}$ and $\omega_{h''k''}$ do not degenerate in a single
spectral line, at least for small $h$ and $k$. This problem is
explained in Section~\ref{sec:mix:dc-mode}.

Other authors write the output frequencies as
$|{\pm}h\omega_l{\pm}k\omega_i|$, with positive $h$ and $k$.  We
recommend to keep the sign of $h$ and $k$.  One reason is that the
positive and negative subscripts of $V_{hk}$ make the spectrum
measurements unambiguously identifiable.  Another reason is that input
phase fluctuations are multiplied by $h$ and $k$, and wrong results
may be obtained discarding the sign.

%================================================================
\subsection{Degenerated Frequency Converter (DC) Mode}\label{sec:mix:dc-mode}
%================================================================
%
The conditions for the mixer to operate in DC mode are the following
\begin{itemize}\item the LO and the RF ports are saturated by sinusoidal signals,
\item the input frequencies are not equal, and the ratio
  $\omega_l/\omega_i$ is equal or close to the the ratio of two small
  integers (say, 5--7 max.),
\item the output is band-passed.
\end{itemize}%
When $\omega_l$ and $\omega_i$ are multiple of a common frequency 
$\omega_0$, thus 
\begin{equation}
\omega_l=p\omega_0\quad\text{and}\quad\omega_i=q\omega_0
\qquad\text{integer}~p{>}0,~q{>}0,~p{\neq}q~~,
\label{eqn:mix:dfc-cond}
\end{equation}
the sum~\req{eqn:mix:sfc-vo-harm} degenerates, and groups of terms
collapse into fewer terms of frequency $n\omega_0$, integer $n$.  The
combined effect of saturation and symmetry produces strong odd-order
harmonics $h\omega_l$ and $k\omega_i$
\begin{align*}
&\omega_l\,{:} &v_{l1}&=V_1\cos(p\omega_0t+\varphi_l)&\qquad
&\omega_i\,{:} &v_{i1}&=V_1\cos(q\omega_0t+\varphi_i)\\ 
&3\omega_l\,{:}&v_{l3}&=V_3\cos(3p\omega_0t+3\varphi_l)&\qquad
&3\omega_i\,{:}&v_{i3}&=V_3\cos(3q\omega_0t+3\varphi_i)\\%[-1ex]
&\cdots        &\cdots       &\qquad\cdots&\qquad
&\cdots        &\cdots       &\qquad\cdots\\[-1ex]
&h\omega_l\,{:}&v_{lh}&=V_h\cos(hp\omega_0t+h\varphi_l)&\qquad
&k\omega_i\,{:}&v_{ik}&=V_k\cos(kq\omega_0t+k\varphi_i)\\%[-1ex]
&\cdots        &\cdots       &\qquad\cdots&\qquad
&\cdots        &\cdots       &\qquad\cdots
\end{align*}
inside the mixer.  After time-domain multiplication, all the cross
products appear, with amplitude $V_{hk}$, frequency $(hp+kq)\omega_0$,
and phase $h\varphi_l+k\varphi_i$.  The generic output term of
frequency $n\omega_0$ derives from the vector sum of all the terms for
which
\begin{equation}
hp+kq=n~~,
\end{equation}
thus
\begin{equation}
  v_n(t)=\sum_{\substack{h,k~\text{pair\,:}\\hp+kq=n}}
  V_{hk}\cos(n\omega_0t+h\varphi_l+k\varphi_i)
\label{eqn:mix:dfc-vn}
\end{equation}
Reality is even more complex than \req{eqn:mix:dfc-vn} because
\begin{itemize}\item some asymmetry is always present, thus even-order harmonics,
\item each term of \req{eqn:mix:dfc-vn} may contain an additional
  constant phase $\varphi_{hk}$,
\item for a given $\omega_l$\,$\omega_i$ pair, several output
  frequencies $n\omega_0$ exist, each one described by
  \req{eqn:mix:dfc-vn}.  Due to nonlinearity, the $v_n(t)$ interact
  with one another.
\end{itemize}%
Fortunately, the amplitudes $V_{hk}$ decrease rapidly with $|hk|$,
therefore the sum \req{eqn:mix:dfc-vn} can be accurately estimated
from a small number of terms, while almost all the difficulty resides
in parameter measurement.  For this reason, there is no point in
devlopping a sophisticated theory, and the few cases of interest can
be anlyzed individually.  The following example is representative of
the reality.

\begin{example}
%----------------------------------------------------------------
The input frequencies are $f_l=5$ MHz and $f_i=10$ MHz, and we select
the output frequency $f_o=5$ MHz with an appropriate bnd-pas filter.
Thus $f_0=5$ MHz, $p=1$, $q=2$, and $n=1$.  The output
signal~\req{eqn:mix:dfc-vn} results from the following terms
\begin{equation*}
\begin{array}{cc|ccc|cl}
hf_l+kf_i=nf_0          &&&hp+kq=n&&&v_n(t)\\\hline
-1{\times}5+1{\times}10=5&&&-1{\times}1{+1}{\times}2=1&&&V_{-1\,1}\cos(\omega_0t{-}\varphi_l{+}\varphi_i)\\
+3{\times}5-1{\times}10=5&&&+3{\times}1{-1}{\times}2=1&&&V_{3\,-1}\cos(\omega_0t{+}3\varphi_l{-}\varphi_i)\\
-5{\times}5+3{\times}10=5&&&-5{\times}1{+3}{\times}2=1&&&V_{-5\,3}\cos(\omega_0t{-}5\varphi_l{+}3\varphi_i)\\
+7{\times}5-3{\times}10=5&&&+7{\times}1{-1}{\times}2=1&&&V_{7\,-3}\cos(\omega_0t{+}7\varphi_l{-}3\varphi_i)\\
-9{\times}5+5{\times}10=5&&&-9{\times}1{+5}{\times}2=1&&&V_{-9\,5}\cos(\omega_0t{+}7\varphi_l{-}3\varphi_i)\\
\cdots                   &&&\cdots                    &&&\cdots
\end{array}%
\end{equation*}%
\end{example}

%----------------------------------------------------------------
\subsection{Phase Amplification Mechanism}\label{sec:mix:phase-ampli}
%----------------------------------------------------------------
\begin{figure}[t]
\centering~~~%     % COMBINED LATEX-PS/PDF FIGURES
	\centering\scalebox{\normalscale}{%  %              % scale
	\input{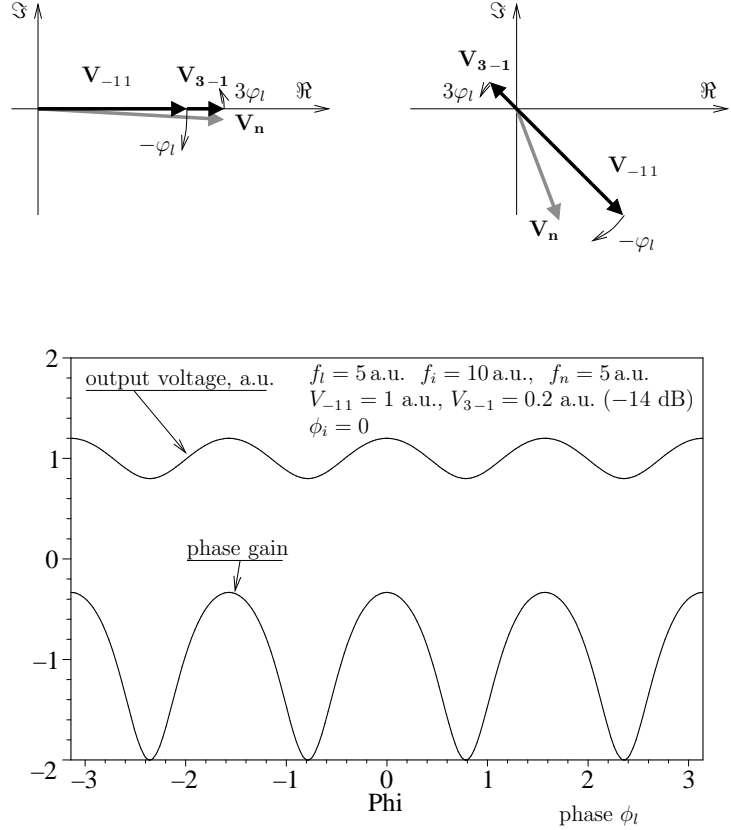}}\par
\caption{Simplified picture of degenerated frequency conversion.
Only $\mathbf{V_{-1\,1}}$ and $\mathbf{V_{3\,{-1}}}$ are taken into account, with $\varphi_i=0$.
Top: phasor representation.
Bottom: output voltage, and phase gain as a function of the static phase $\varphi_l$.}
\label{fig:mix-phasors}
\end{figure}

Introducing the phasor (Fresnel vector) representation\footnote{In this section we use uppercase boldface for phase vectors, as in $\mathbf{V}=Ve^{j\varphi}$.  $V$ is the rms voltage.}
Eq.~\req{eqn:mix:dfc-vn} becomes $\mathbf{V_n}=\sum\mathbf{V_{hk}}$,
thus
\begin{equation}
\frac{1}{\sqrt{2}}\,V_n\,e^{j\varphi_n}=
\sum_{\substack{h,k~\text{pair\,:}\\hp+kq=n}}
\frac{1}{\sqrt{2}}\,V_{hk}\,e^{j\varphi_{hk}}
\qquad\text{with}\;\varphi_{hk}=h\varphi_l+k\varphi_i~~.
\label{eqn:mix:dfc-vn-vec}
\end{equation}
Both $V_n$ and $\varphi_n$ are function of $\varphi_l$ and
$\varphi_i$, thus function of the phase relationship between the two
inputs.  Let $\phi$ the fluctuation of the static phase $\varphi$.  The
output phase fluctuation is
\begin{equation}
\phi_n=
\frac{\partial\varphi_n}{\partial\varphi_l}\,\phi_l +
\frac{\partial\varphi_n}{\partial\varphi_i}\,\phi_i~~,
\label{eqn:mix:dfc-phasegain}
\end{equation}
where the derivatives are evaluated in the static working point.  
There follows that the input phase fluctuations $\varphi_l$ and $\varphi_i$ are amplified or attenuated (gain lower than one) by the mixer.
The phase gain/attenuation mechanism is a consequence of degeneracy.  The effect on phase noise was discovered studying the regenerative frequency dividers \cite{rubiola92im}.

Figure~\ref{fig:mix-phasors} shows a simplified example in which a 5~MHz
signal is obtained by mixing a 5~MHz and a 10~MHz, accounting only for two modes ($10-5$ and $3{\times}5-10$).  For $\varphi_l=0$, the
vectors are in phase, and the amplitude is at its maximum.  A small
negative $\varphi_n$ results from $\mathbf{V_{-1\,1}}$ and
$\mathbf{V_{3\,{-1}}}$ pulling in opposite directions.  A phase fluctuation is therefore attenuated.  For
$\varphi_l=\pi/4\simeq0.785$, the vectors are opposite, and the
amplitude is at its minimum.  The combined effect of
$\mathbf{V_{-1\,1}}$ and $\mathbf{V_{3\,{-1}}}$ yields a large
negative $\varphi_n$.  With $V_{3\,{-1}}/V_{-1\,1}=0.2$ ($-14$ dB),
the phase gain $\partial\varphi_n/\partial\varphi_l$ spans from
$-0.33$ and $2$, while it would be $-1$ (constant) if only the $-1,1$
mode was present.

The experimentalist not aware of degeneracy may obtain disappointing
results when low-order harmonics are present, as in the above example.
The deliberate exploitation of degeneracy to manage phase noise is one
of the most exhotic uses of the mixer.

\begin{figure}[t]
\centering%     % COMBINED LATEX-PS/PDF FIGURES
	\centering\scalebox{\normalscale}{%  %              % scale
	\input{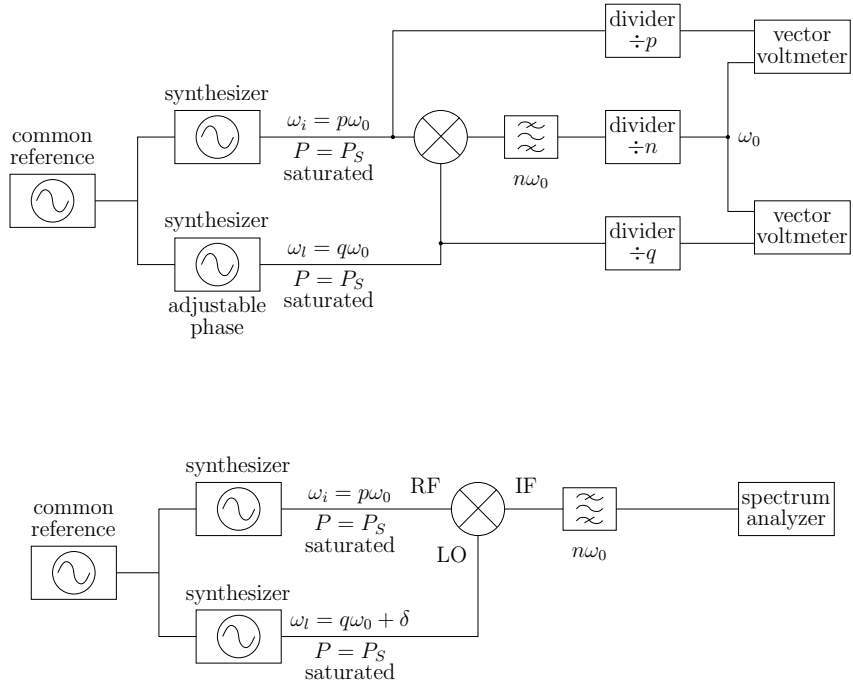}}
\caption{Parameter masurement of a degenerated frequency converter.}
\label{fig:mix-dfc-meas}
\end{figure}
\begin{figure}[t]
\centering%     % COMBINED LATEX-PS/PDF FIGURES
	\centering\scalebox{\smallscale}{%  %              % scale
	\input{\includefigpathmix-phase-gain-spectrum.pstex_t}}
\caption{Amplitude, phase, and phase gain in a 
  degenerated frequency converter.}
\label{fig:mix-phase-gain-spectrum}
\end{figure}
\begin{figure}[t]
\centering%     % COMBINED LATEX-PS/PDF FIGURES
	\centering\scalebox{\LARGEscale}{%  %              % scale
	\input{\includefigpathmix-phase-gain.pstex_t}}
\caption{Amplitude, phase, and phase gain in a 
  degenerated frequency converter.}
\label{fig:mix-phase-gain}
\end{figure}
%----------------------------------------------------------------
\paragraph{Parameter Measurement.}
%----------------------------------------------------------------
There are two simple ways to measure the parameters of a degenerated
frequency converter (Fig.~\ref{fig:mix-dfc-meas}).

The first method is the separate measurement of the coefficients
$V_{hk}$ of Eq.~\req{eqn:mix:dfc-vn} by means of a spectrum analyzer.
One input signal is set at a frequency $\delta$ off the nominal
frequency $\omega_l$ (or $\omega_i$).  In this condition degeneracy is
broken, and all the terms of Eq.~\req{eqn:mix:dfc-vn} are visible as
separate frequencies.  The offset $\delta$ must be large enough to enable the accurate measurement of all the spectral lines with a spectrum
analyzer, but small enough not to affect the mixer operation.  Values
of 10--50 kHz are useful in the HF/UHF bands, and up to 1 MHz at
higher frequencies.  Figure~\ref{fig:mix-phase-gain-spectrum} provides an example.
This method is simple and provides insight.
On the other hand, it is not very accurate because it hides the phase errors $\varphi_{hk}$ that may be present in each term.

The second method consists of the direct measurement of $\mathbf{V_n}$
[Eq.~\req{eqn:mix:dfc-vn-vec}] as a function of the input phase,
$\varphi_l$ or $\varphi_i$, by means of a vector voltmeter.  This gives
amplitude and phase, from which the phase gain is derived.  For the
measurement to be possible, the three signals must be converted to the
same frequency $\omega_0$ with approprate dividers.  Of course, the
mixer must be measured in the same conditions (RF and LO power) of the
final application.  While one vector voltmter is sufficient, it is
better to use two vector voltmters because the measurement accounts
for the reflected waves in the specific circuit.  In some cases good results are
obtained with resistive power splitters located close to the mixer
because these splitters are not directional.  Interestingly, most
frequency synthesizers can be adjusted in phase even if this feature
is not explicitely provided.  The trick consists of misaligning the
internal quartz oscillator when the instrument is locked to an
external frequency reference.  If the internal phase locked loop does
not contain an integrator, the misalignamet turns into a phase shift,
to be determined a posteriori.  The drawback of the direct measurement
method is that it requires up to two vector voltmeters, two frequency
synthesizers and three frequency dividers.  In the general case, the
dividers can not be replaced with commercial synthesizers because a
synthesizer generally accepts only a small set of round input
frequencies (5~MHz or 10~MHz).
Figure~\ref{fig:mix-phase-gain} shows an example of direct measurement,
compared to the calculated values, based on the first method.

%===========================================
\subsection{Phase Detector (PD) Mode}\label{sec:mix:pd-mode}
%===========================================
The mixer works as a phase detector in the following conditions
\begin{itemize}\item the LO and the RF ports are saturated by sinusoidal signals of
  the same frequency $\omega_0$, about in quadrature,
\item the output is low-passed.
\end{itemize}%
\begin{figure}[t]
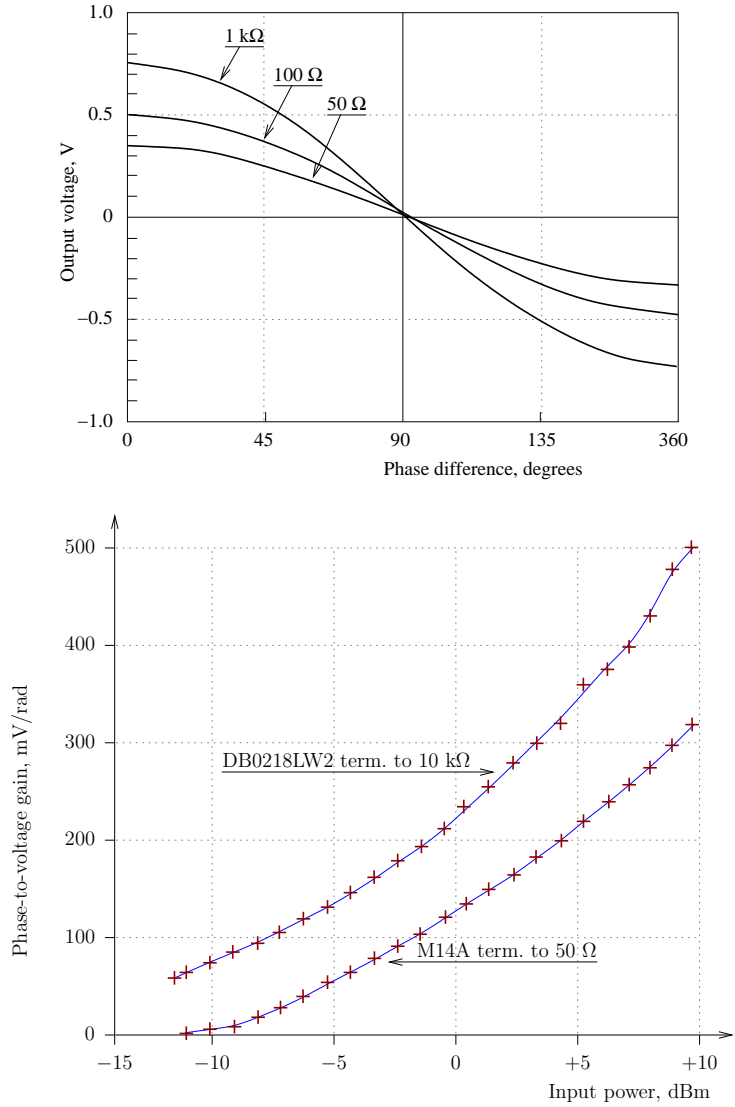

\centering%     % COMBINED LATEX-PS/PDF FIGURES
	\centering\scalebox{\smallscale}{%  %              % scale
	\input{\includefigpathmix-vphi.pstex_t}}\\[1em]
\centering%     % COMBINED LATEX-PS/PDF FIGURES
	\centering\scalebox{\smallscale}{%  %              % scale
	\input{\includefigpathmix-vphi-measured.pstex_t}}
\caption{Example of phase detector characteristics:
  output voltage as a function of $\phi$ (data are from a handbook Macom) and
  phase-to-voltage gain as a function of power (measured).}
\label{fig:mix-vphi}
\label{fig:mix-vphi-measured}
\end{figure}
The product of such input signals is
\begin{equation}
\cos\Bigl(\omega_0t+\varphi\Bigr)\,
\cos\Bigl(\omega_0t-\frac{\pi}{2}\Bigr) =
\frac{1}{2}\,\sin\Bigl(2\omega_0t+\varphi\Bigr) -
\frac{1}{2}\,\sin\varphi~~,
\label{eqn:mix:pd-base}
\end{equation}
from which one obtains a sinusoid of frequency $2\omega_0$, and a dc
term $-\frac{1}{2}\sin\varphi$ that is equal to $-\frac{1}{2}\varphi$
for small $\varphi$.  The output signal of an actual mixer is a
distorted sinusoid of frequency $2\omega_0$ plus a dc term, which can
be approximated by
\begin{equation}
v_o(t)=V_2\sin\bigl(2\omega_0t+\varphi\bigr) - V_0\sin\varphi~~.
\label{eqn:mix:pd-sat}
\end{equation}
$V_2$ and $V_0$ are experimental parameters that depend on the
specific mixer and on power.  Due to saturation, the maximum of
$|v_o(t)|$ is about independent of $\phi$, hence $V_2$ decreases as
the absolute value of the dc term increases.

Using the $2\omega_0$ output signal to double the input frequency is a
poor choice because (i) the quadrature condition can only be obtained
in a limited bandwidth, (ii) the IF circuit is usually designed for
frequencies lower than the RF and LO.  A better choice is to use a
reversed mode.

When the PD mode is used close to the quadrature conditions, the
deviation of dc response from $\sin\varphi$ can be ignored.  After
low-pass filtering, the output signal is\footnote{The
  phase-to-voltage gain is also written as $k_\phi$ (with the
  alternate shape of $\phi$) because it is used with the small
  fluctuations $\phi$.}
\begin{equation}
v_o= - k_\varphi\varphi + V\p{os}~~,
\label{eqn:mix:pd-real}
\end{equation}
where $k_\varphi$ is the phase-to-voltage gain [the same as $V_0$ in
Eq.~\req{eqn:mix:pd-sat}], and $V\p{os}$ is the dc offset that derives
from asymmetry.  Figure~\ref{fig:mix-vphi} shows an example of phase
detector charactaristics.  The IF output can be loaded to a high
resistance in order to increase the gain $k_\varphi$.

It is often convenient to set the input phase for zero dc output, which
compensate for $V\p{os}$.  This condition occurs at some random---yet
constant---phase a few degrees off the quadrature conditions, in a
range where the mixer characteristics are virtually unaffected.

Due to diode asymmetry, the input power affects $V\p{os}$.  Exploiting
the asymmetry of the entire $v(i)$ law of the diodes, it is often
possible to null the output response to the fluctuation of the input
power, therefore to make the mixer insensitive to amplitude
modulation.  This involves setting the phase between the inputs to an
appropriate value, to be determined experimentally.  In our
experience, the major problem is that there are distinct AM
sensitivities
\begin{equation}
\frac{dv_o}{dP_l},\qquad
\frac{dv_o}{dP_i},\qquad 
\frac{dv_o}{d(P_l+P_i)}~~, 
\end{equation}
and that nulling one of them is not beneficial to the other two.  In
some cases the nulls occurr within some 5\degrees\ from the
quadrarure, in other cases farther, where the side effects of the
offset are detrimental.

%============================================
\section{Reversed Modes}\label{sec:mix:reversed-modes}
%============================================
%
\begin{figure}[t]
\centering%     % COMBINED LATEX-PS/PDF FIGURES
	\centering\scalebox{\normalscale}{%  %              % scale
	\input{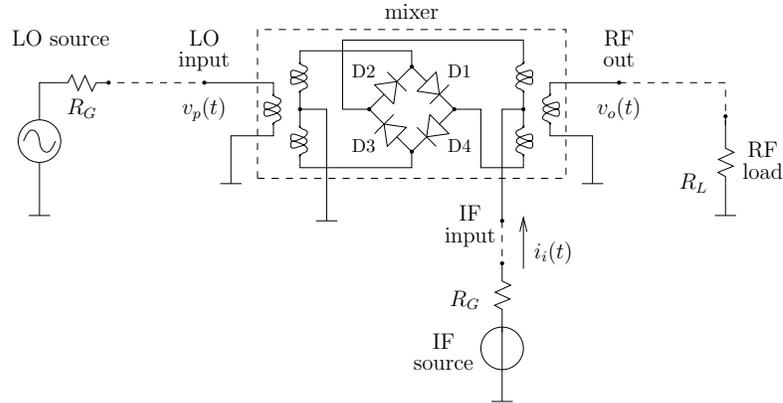}}
\caption{Reversed-mode modulator.}
\label{fig:mix-modul}
\end{figure}
The mixer can be reversed taking the IF port as the input and the RF
port as the output (Fig.~\ref{fig:mix-modul}).  The LO signal makes
the diodes switch, exactly as in the normal modes.  The major
difference versus the normal modes is the coupling bandwidth: the
output is now ac-coupled via the RF balun, while the input is in most cases
dc-coupled.  When impedance-matching is not needed, the IF input can
be driven with a current source.

%============================================
\subsection{Linear Modulator (LM)}\label{sec:mix:lm-mode}
\begin{figure}[t]
\centering%     % COMBINED LATEX-PS/PDF FIGURES
	\centering\scalebox{\normalscale}{%  %              % scale
	\input{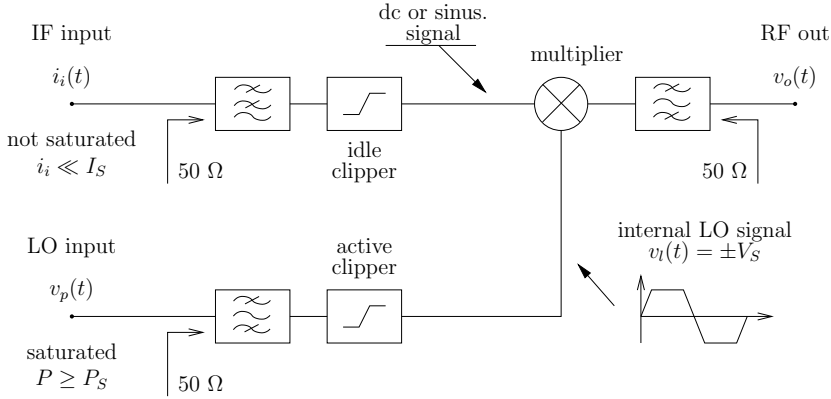}}
\caption{Reverse-mode model of a mixer.}
\label{fig:mix-rev-model}
\end{figure}

%============================================
The mixer works as a LM in the following conditions
\begin{itemize}\item the LO port is saturated by a sinusoidal signal,
\item a near-dc signal is present at the IF input,
\item the IF input current is lower than the saturation
  current\footnote{The mixer saturation current, which can be of some
    mA, should not be mistaken for the diode reverse saturation
    current.  The latter can be in the range from $10^{-15}$ A to $10^{-15}$ A.}~$I_S$.
\end{itemize}%
As usual, the LO pump forces the diodes to switch.  At zero input
current, due to symmetry, no signal is present at the RF output.  When
a positive current $i_i$ is present, the resistance of D2 and D4
averaged over the period decreases, and the conduction angle of D2 and
D4 increases.  The average resistance of D1 and D3 increases, and
their conduction angle decreases.  Therefore, a small voltage $v_o(t)$
appears at the RF output, of amplitude proportional to $i_i$, in phase
with $v_p(t)$.  Similarly, a negative $i_i$ produces an output voltage
proportional to $i_i$, of phase opposite to $v_p(t)$.  The mixer can
be represented as the system of Fig.~\ref{fig:mix-rev-model}, which is
similar to the LC model (Fig.~\ref{fig:mix-lc-model}) but for the
input-output filters.
The internal saturated LO signal can be approximated with a sinusoid
$v_l(t)=V_L\cos\omega_lt$, [Eq.~\req{eqn:mix:lo}], or expanded as
Eq.~\req{eqn:mix:multih-lo-1}.  Strictly, $V_L$ can not be derived
from the reverse loss, which is not documented.  Reciprocity should
not given for granted.  Nonetheless, measuring some mixers we found
that the `conventional' (forward) SSB loss $\ell$ and
Eq.~\req{eqn:mix:equiv-lo-v} provide useful approximation of reverse
behavior.  Thus, the mixer operates as a linear modulator described by
\begin{align}
v_o(t) &= \frac1U\,v_i(t)\,v_l(t)\\[1ex]
       &= \frac1U\,v_i(t)\,V_L\cos\omega_lt~~.
\label{eqn:mix:rev-mod-dc}
\end{align}

\begin{figure}[t]
\centering%     % COMBINED LATEX-PS/PDF FIGURES
	\centering\scalebox{\smallscale}{%  %              % scale
	\input{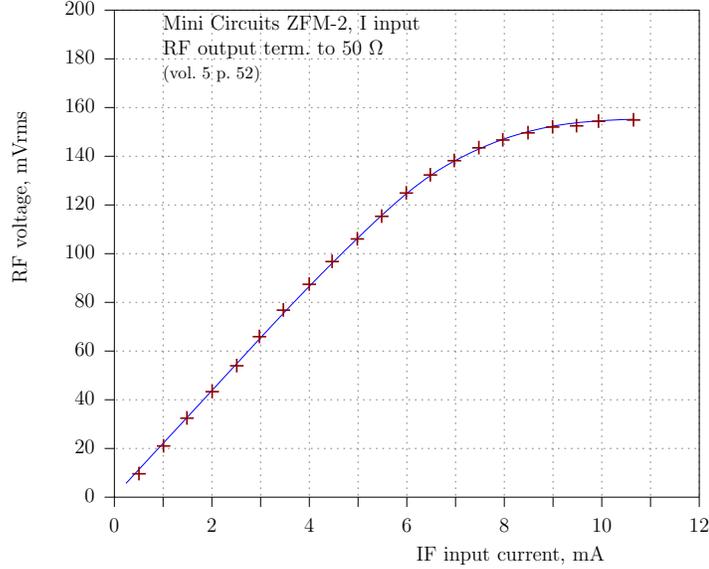}}
\caption{Gain of a mixer used ad a modulator.
  Input is driven with a current source.  Output is terminated to
  50~\ohm.}
\label{fig:mix-iq-mod-gain}
\end{figure}

\begin{example}
%----------------------------------------------------------------
The LO signal of a mixer (Mini-Circuits ZFM-2) is a sinusoid of
frequency $f_l=100$ MHz and power $P=5$ mW (7~dBm).  In such
conditions the nominal SSB loss is $\ell=2$ (6~dB).  By virtue of
Eq.~\req{eqn:mix:equiv-lo-v}, $V_L=1$~V\@.  When the input current is
$i_i=2$~mA dc, the input voltage is $v_i=R_0i_i=100$~mV with
$R_0=50$~\ohm.  After Eq.~\req{eqn:mix:rev-mod-dc}, we expect an output
signal of 100 mV peak, thus 71 mV rms.  This is close to the measured
value of 75 mV\@.  The latter is obtained fitting the the low-current
experimental data of Fig.~\ref{fig:mix-iq-mod-gain}.  Beyond $i_i=3$~mA,
the mixer lives gradually the linear behavior, and saturates at some
230 mV rms of output signal, when $i_i\approx12$~mA dc.  Similar
results were obtained testing other mixers.
\end{example}

%============================================
\subsection{Reverse Linear Converter (RLC)}\label{sec:mix:rlc-mode}
%============================================
%
The mixer works as a RLC in the following conditions
\begin{itemize}\item the LO port is saturated by a sinusoidal signal,
\item a small narrowband signal is present at the IF input, which is
  not saturated,
\item LO and the IF separated in the frequency domain,
\item an optional filter selects one of the beat products.
\end{itemize}%
This mode is similar to the LM mode.  Letting
$v_i(t)=A_i(t)\cos[\omega_i(t)+\varphi_i(t)]$ the input, the output
signal is
\begin{align}
v_o(t) & = \frac{1}{U}~v_i(t)\,v_l(t) \\[0.5ex]
       & = \frac{1}{U}~A_i(t)\cos\bigl[\omega_it+\varphi_i(t)\bigr]%
           ~~V_L\cos(\omega_lt) \\[0.5ex]
       & = \frac{1}{2U}\,V_LA_i(t)\:\Bigl\{
           \cos\bigl[(\omega_l-\omega_i)t-\varphi_i(t)\bigr]+
           \cos\bigl[(\omega_l+\omega_i)t+\varphi_i(t)\bigr]\Bigr\}~~.
\label{eqn:mix:rev-mod-ac}
\end{align}
The model of Fig.~\ref{fig:mix-rev-model} still holds, and the
internal LO amplitude $V_L$ can be estimated using
Eq.~\req{eqn:mix:equiv-lo-v} and the `conventional' SSB loss $\ell$.

If an external bandpass filter, not shown in Fig.~\ref{fig:mix-rev-model},
is present, the output signal is
\begin{align}
v_o(t) &= \frac{1}{2U}\,V_LA_i(t)\,
          \cos\bigl[(\omega_l-\omega_i)t-\varphi_i(t)\bigr]
          \qquad\text{LSB,}\qquad\qquad\text{or}\\[1ex]
v_o(t) &= \frac{1}{2U}\,V_LA_i(t)\,
          \cos\bigl[(\omega_l+\omega_i)t+\varphi_i(t)\bigr]
          \qquad\text{USB}~~,
\end{align}
under the obvious condition that the signal bandwidth fits into the
filter passband.

%===========================================
\subsection{Digital Modulator (DM) Mode}\label{sec:mix:dm-mode}
%===========================================
The mixer works as a DM in the following conditions
\begin{itemize}\item the LO port is saturated by a sinusoidal signal,
\item a large near-dc current is present at the IF input, which is
  saturated,
\item the RF output is bandpassed.
\end{itemize}%
Let $v_p=V_P\cos\omega_lt$ the LO input signal, $i_i={\pm}I_i$ the IF
input current, and $V_O$ the saturated output amplitude.  The output
signal is
\begin{equation}
v_o(t) = \mbox{sgn}(i_i)\:V_O\cos\omega_lt~~,
\label{eqn:mix:dm-out} 
\end{equation}
where $\mbox{sgn}(\cdot)$ is the signum function.
Equation~\req{eqn:mix:dm-out} represents a BPSK (binary phase shift keying)
modulation driven by the input current $i_i$.

%================================================================
\subsection{Reverse Saturated Converter (RSC) Mode}\label{sec:mix:rsc-mode}
%================================================================
%
The mixer works in the RSC mode under the following conditions 
\begin{itemize}\item the LO and the IF ports are saturated by sinusoidal signals,
\item the input frequencies are not equal, and the ratio
  $\omega_l/\omega_i$ is not too close to the the ratio of two small
  integers (say, 5-7 max.),
\item the output is band-passed.
\end{itemize}%
The RSC mode is similar to the SC mode, for the explanations given in
Section~\ref{sec:mix:sc-mode} also apply to the RSC mode.  The only
difference between SC and RSC is the input and output bandwidth,
because IF and RF are interchanged.

%=============================================
\subsection{Reverse Degenerated Converter (RDC) Mode}\label{sec:mix:rdc-mode}
%=============================================
The mixer works in the RDC mode when
\begin{itemize}\item the LO and the IF ports are saturated by sinusoidal signals,
\item the input frequencies are equal, or the ratio
  $\omega_l/\omega_i$ is equal or close to the the ratio of two small
  integers (say, no more than 5--7),
\item the output is band-passed.
\end{itemize}%
The RDC mode is similar to the DC mode (Section~\ref{sec:mix:sc-mode})
but for the trivial difference in the input and output bandwidth, as the roles of 
IF and RF are interchanged.  The output signal results from the 
vector addition of several beat signals, each one with its own
phase and amplitude.   

It is to be made clear that when two equal input frequencies ($\omega_i=\omega_l=\omega_0$) are sent to the input, the reverse mode differs significantly from the normal mode.  In the DC mode, this condition would turn the degenerated converter mode into the phase-detector mode.  But in the reversed modes no dc output is permitted because the RF port is ac coupled.  Of course, a large $2\omega_0$ signal is always present at the RF output, resulting from the vector addition of
several signals, which makes the RDC mode an efficient frequency doubler.

\begin{example}
%----------------------------------------------------------------
The input frequencies are $f_l=f_i=5$ MHz, and we select the output
$f_o=10$ MHz.  Thus $f_0=5$ MHz, $p=1$, $q=1$, and $n=2$.  The output
signal~[Eq.~\req{eqn:mix:dfc-vn}] results from the follwoing terms
\begin{equation*}
\begin{array}{cc|ccc|cl}
hf_l+kf_i=nf_0           &&&hp+kq=n&&&v_n(t)\\\hline
+1{\times}5+1{\times}5=10&&&+1{\times}1{+1}{\times}1=2&&&V_{1\,1}\cos(\omega_0t{+}\varphi_l{+}\varphi_i)\\
+3{\times}5-1{\times}5=10&&&+3{\times}1{-1}{\times}1=2&&&V_{3\,{-1}}\cos(\omega_0t{+}3\varphi_l{-}\varphi_i)\\
-1{\times}5+3{\times}5=10&&&-1{\times}1{+3}{\times}1=2&&&V_{-1\,3}\cos(\omega_0t{-}\varphi_l{+}3\varphi_i)\\
+5{\times}5-3{\times}5=10&&&+5{\times}1{-3}{\times}1=2&&&V_{5\,-{3}}\cos(\omega_0t{+}5\varphi_l{-}3\varphi_i)\\
-3{\times}5+5{\times}5=10&&&-3{\times}1{+5}{\times}1=2&&&V_{-3\,5}\cos(\omega_0t{-}3\varphi_l{+}5\varphi_i)\\
\cdots                   &&&\cdots                    &&&\cdots
\end{array}
\end{equation*}
\end{example}

%================================================================
\section{Special Mixers and I-Q Mixers}\label{sec:mix:specials-iqs}
%================================================================
%
\begin{figure}[t]
\centering%     % COMBINED LATEX-PS/PDF FIGURES
	\centering\scalebox{\normalscale}{%  %              % scale
	\input{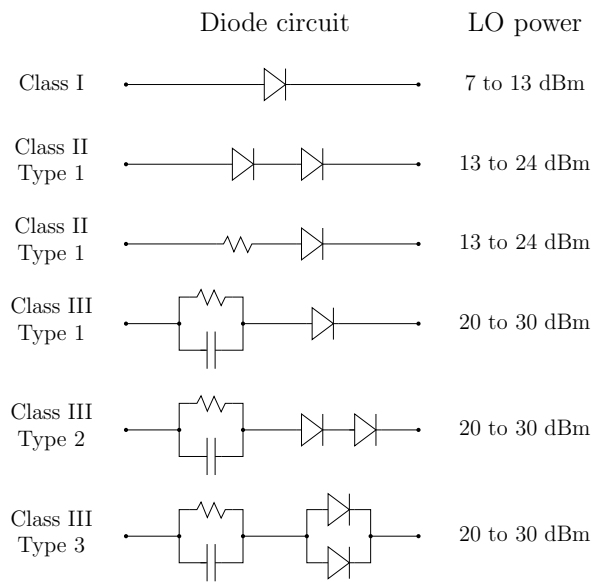}}
\caption{Diode assemblies of high linearity mixers.}
\label{fig:mix-multidiode}
\end{figure}

\paragraph{Phase Detector.}  
%----------------------------------------------------------------
Some mixers are explicitely designed to operate in the phase detector
mode.  In some cases such devices are actually general-purpose mixers \emph{documented}
for phase detector operation.  Often the IF output impedance is larger than 50 \ohm, typically 500 \ohm.  The main advantage of this higher impedance is a lower residual white noise of the system.  In fact, the output preamplifier can hardly be noise-matched to an input
resistance lower than a few hundreds Ohms.  The IF bandwidth reduction that results from the increased output impedance is not relevant in practice.  The residual flicker, which is the most relevant parameter for a number of measurements, is usually not documented\footnote{I never come across a phase detector whose residual flicker is documented.}.

\paragraph{Analog Modulator / Variable Attenuator.}  
%----------------------------------------------------------------
A mixer can be designed and \emph{documented} to be used in a reverse mode as an analog modulator (See Sec.~\ref{sec:mix:lm-mode}).  The fancy name ``variable attenuator'' is sometimes used.  Yet, the mixer operation is more general than that of a simple attenuator because the mixer input current can be either positive or negative, and the output signal changes sign when the input current is negative.

\paragraph{BPSK Modulator.}  
%----------------------------------------------------------------
The BPSK modulator differs from the analog modulator in that the IF input is saturated (See Sec.~\ref{sec:mix:dm-mode}).  Once again, the device may differ from a general-purpose mixer mostly in the documentation.

\paragraph{High Linearity Mixers.}  
%----------------------------------------------------------------
In some cases low intermodulation performance must be achieved at any
cost.  Special mixers are used, based on a ring in which the diodes
are replaced with the more complex elements shown in
Fig.~\ref{fig:mix-multidiode} (classes~I-III).  High linearity is
achieved by forcing the diodes to switch abruptly in the presence of a
large pump signal.  These mixers, as compared to the single-diode
ones, need large LO power, up to 1 W, and show higher loss.

\begin{figure}[t]
\centering%     % COMBINED LATEX-PS/PDF FIGURES
	\centering\scalebox{\normalscale}{%  %              % scale
	\input{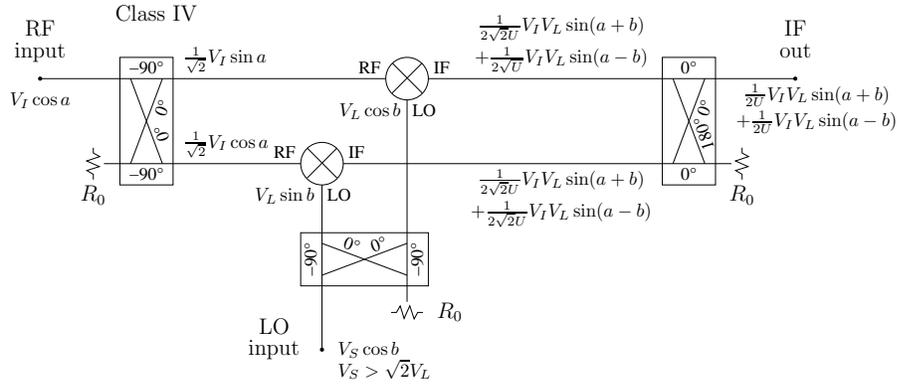}}
\caption{Improved impedance-matching mixer.}
\label{fig:mix-improved-z-match}
\end{figure}

\paragraph{Improved Impedance-Matching Mixers.}  
%----------------------------------------------------------------
The 90\degrees\ hybrid junction, used as a power splitter, 
has the useful property that the input (output) is always impedance
matched when the isolation port is correctly terminated and the two
outputs (inputs) are loaded with equal impedances.  This property is
exploited joining two equal double-balanced mixers to form the
improved mixer of Fig.~\ref{fig:mix-improved-z-match} (Class~IV mixer).  Other
schemes are possible, based on the same idea.

\paragraph{Double-Double-Balanced Mixers.}  
%----------------------------------------------------------------
\begin{figure}[t]
\centering%     % COMBINED LATEX-PS/PDF FIGURES
	\centering\scalebox{\normalscale}{%  %              % scale
	\input{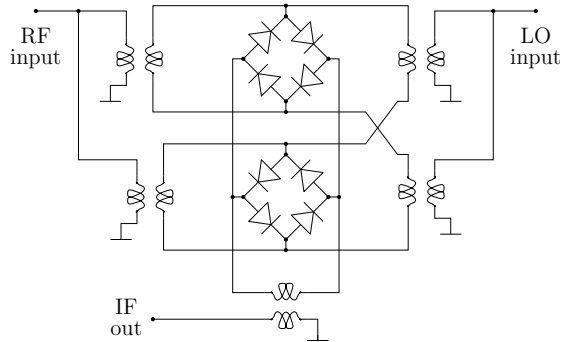}}
\caption{Double-double-balanced mixer.}
\label{fig:mix-ddbm}
\end{figure}
The double-double-balanced mixer (Figure~\ref{fig:mix-ddbm}) shows high
1 dB compression point, thus high dynamic range and low distortion,
and high isolation.  This device is sometimes called \emph{triple
  balanced mixer} because it is balanced at the three ports.  Other
schemes are possible.

\paragraph{Image-Rejection Mixer.}  
%----------------------------------------------------------------
%
\begin{figure}[t]
\centering%     % COMBINED LATEX-PS/PDF FIGURES
	\centering\scalebox{\normalscale}{%  %              % scale
	\input{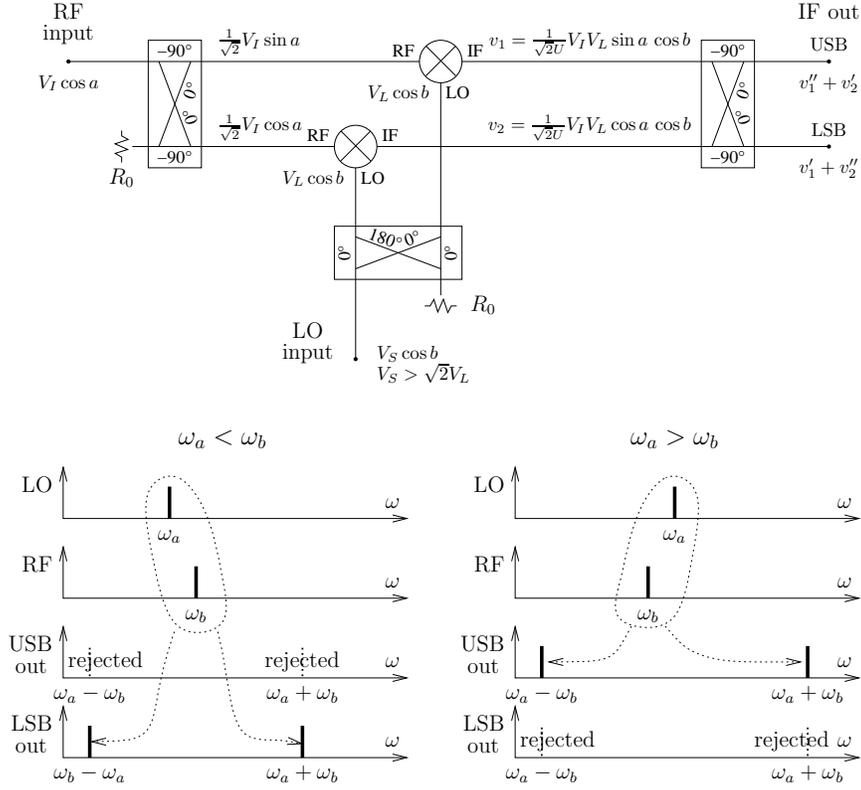}}
\caption{Image-rejection mixer.}
\label{fig:mix-img-rej}
\end{figure}
Let us go back to the frequency conversion system of
Fig.~\ref{fig:mix-lc-image}, in which the LSB and the USB are
converted into the same IF frequency $\omega_b$.  The scheme of
Fig.~\ref{fig:mix-img-rej} divides the IF components, enabling the
selection of the LSB or the USB input (RF) signal.  
 
Let us for short $a=\omega_it$ and $b=\omega_lt$ the instantaneous phase of the RF and LO signal.  The converted signals, at the IF output of the mixers are 
\begin{align*}
v_1&=\frac{1}{\sqrt{2}U} V_IV_L\, \sin a  \, \cos b\\[0.5ex]
v_2&=\frac{1}{\sqrt{2}U} V_IV_L\, \cos a \, \cos b~~,
\end{align*}
thus
\begin{align*}
v_1&=\frac{1}{2\sqrt{2}U} V_IV_L \Bigl[\sin(a-b)+\sin(a+b)\Bigr] \\[0.5ex] 
v_2&=\frac{1}{2\sqrt{2}U} V_IV_L \Bigl[\cos(a-b)+\cos(a+b)\Bigr]~~.
\end{align*}
The path of the hybrid junction labeled  `$-90^\circ$' turns the phase of the positive-frequency signals by $-90^\circ$, and the phase of the negative-frequencies signal by $+90^\circ$. The
rotated signals are
\begin{align*} 
v''_1  &=\begin{cases}
          \frac{1}{4U} V_IV_L\bigl[-\cos(a-b)-\cos(a+b)\bigr]    &a{>}b\\[0.5ex]
          \frac{1}{4U} V_IV_L\bigl[+\cos(a-b)+\cos(a+b)\bigr]  &a{<}b
          \end{cases}\\[2ex]
v''_2  &=\begin{cases}
          \frac{1}{4U} V_IV_L \bigl[+\sin(a-b)+\sin(a+b)\bigr]  &a{>}b\\[0.5ex] 
          \frac{1}{4U} V_IV_L \bigl[-\sin(a-b)-\sin(a+b)\bigr]    &a{<}b
          \end{cases}
\end{align*}
which also account for a factor $1/\sqrt{2}$ due to energy conservation.  The non-rotated signals are
\begin{align*}
v'_1&=\frac{1}{4U} V_IV_L \bigl[\sin(a-b)+\sin(a+b)\bigr] \\[0.5ex] 
v'_2&=\frac{1}{4U} V_IV_L \bigl[\cos(a-b)+\cos(a+b)\bigr]~~.
\end{align*}
The output signals are
\begin{align} 
v_\text{USB}=v''_1+v'_2&=\begin{cases} 
          \frac{1}{4U} V_IV_L \bigl[\sin(a-b)+\sin(a+b)\bigr]
          & a{>}b~~\text{\footnotesize (USB taken in)}\\[0.5ex] 
          0                      
          & a{<}b~~\text{\footnotesize (LSB rejected)}
          \end{cases} \\[2ex] 
v_\text{LSB}=v'_1+v''_2&=\begin{cases} 
          0                  
          &a{>}b~~\text{\footnotesize (USB rejected)}\\[0.5ex]
          \frac{1}{4U} V_IV_L \bigl[\cos(a-b)+\cos(a+b)\bigr]
          &a{<}b~~\text{\footnotesize (LSB taken in)} 
          \end{cases}
\end{align}

The unwanted sideband is never cancelled completely.  A rejection of 20 dB is common in practice.
The main reason to prefer the image-rejection mixer to a (simple) mixer is noise.  Let us assume that the LO frequency $\omega_l$ and the IF center frequency $\omega_\text{IF}$ are given.
The mixer converts both $|\omega_l-\omega_\text{IF}|$ and $|\omega_l+\omega_\text{IF}|$ to $\omega_\text{IF}$, while the image-rejection mixer converts only one of these channels.  Yet, the noise of the electronic circuits is present at both frequencies.

\begin{example}
The IF filter of a FM receiver has a bandwidth of 300 kHz centered at 10.7 MHz.  In order to receive a channel at 91 MHz, we tune the local oscillator to 101.7 MHz ($101.7-10.7=91$).  A mixer down-convert to IF two channels, the desired one (91 MHz) and the image frequency at 122.4 MHz ($101.7+10.7=122.4$).  In the best case, only noise is present at the image frequency (122.4 MHz), which is taken in by the mixer, yet not by the image-rejection mixer.
\end{example}

\paragraph{SSB Modulator.}  
%----------------------------------------------------------------
\begin{figure}[t]
\centering%     % COMBINED LATEX-PS/PDF FIGURES
	\centering\scalebox{\normalscale}{%  %              % scale
	\input{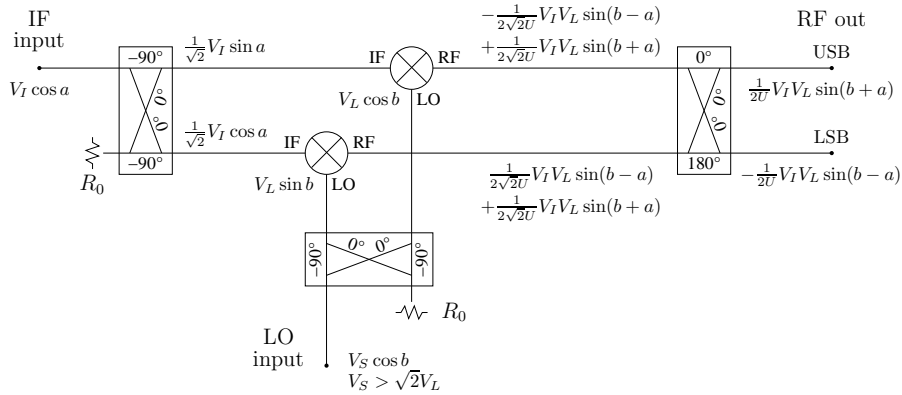}}
\caption{SSB modulator.}
\label{fig:mix-ssb-mod}
\end{figure}
\begin{figure}[t]
\centering%     % COMBINED LATEX-PS/PDF FIGURES
	\centering\scalebox{\smallscale}{%  %              % scale
	\input{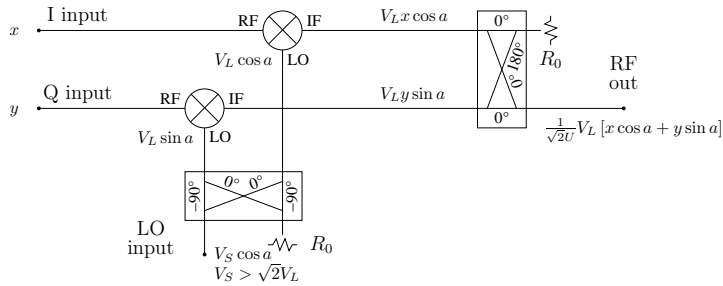}}
\caption{I-Q modulators.}
\label{fig:mix-iq-modulators}
\end{figure}
The SSB modulator (Fig.~\ref{fig:mix-ssb-mod}) is a different arrangement of the same blocks used in the image-rejection mixer.  The main purpose of this device is to modulate a carrier by adding only one sideband, either LSB or USB\@.  All explanations are given on the scheme, in Fig.~\ref{fig:mix-ssb-mod}.

\paragraph{I-Q Detectors and Modulators.}  
%----------------------------------------------------------------
\begin{figure}[ht]
\centering%     % COMBINED LATEX-PS/PDF FIGURES
	\centering\scalebox{\smallscale}{%  %              % scale
	\input{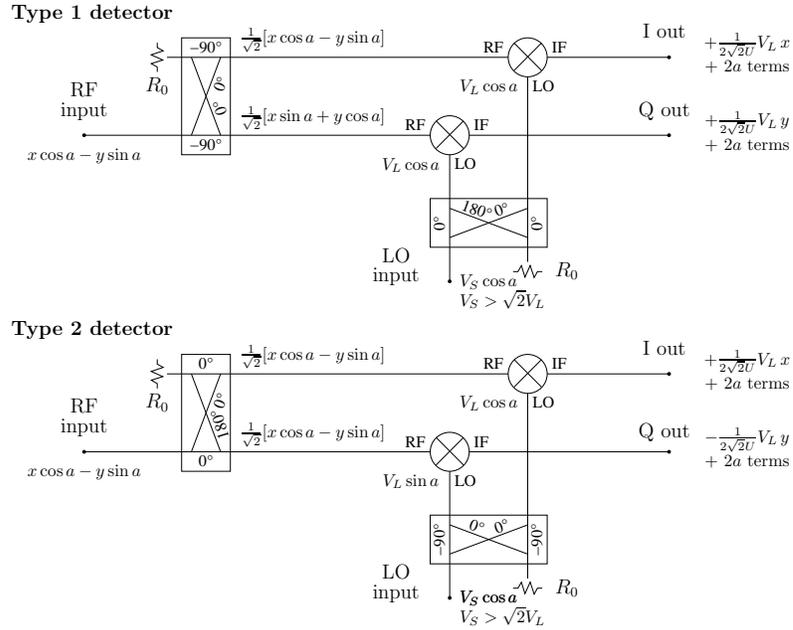}}
\caption{I-Q detectors.}
\label{fig:mix-iq-detectors}
\end{figure}
The two-axis synchronous detector introduced in Section~\ref{ssec:mix:sd-mode} is commercially available in (at least) two practical implementations, shown in Fig.~\ref{fig:mix-iq-detectors}.
Of course, the conversion loss is increased by the loss of the input power splitter,
which is of 3--4 dB\@.  For the same reason, the required LO power is increased by 3--4 dB\@.  
The I-Q mixer can be reversed, operating as a modulator, as the simple mixer did (Sec.~\ref{sec:mix:lm-mode}).  A number of I-Q modulators are available off the shelf, shown in Fig.~\ref{fig:mix-iq-modulators}.  
Other configurations of I-Q detector/modulator are possible, with similar characteristics.

\begin{figure}[t]
%     % COMBINED LATEX-PS/PDF FIGURES
	\centering\scalebox{\smallscale}{%  %              % scale
	\input{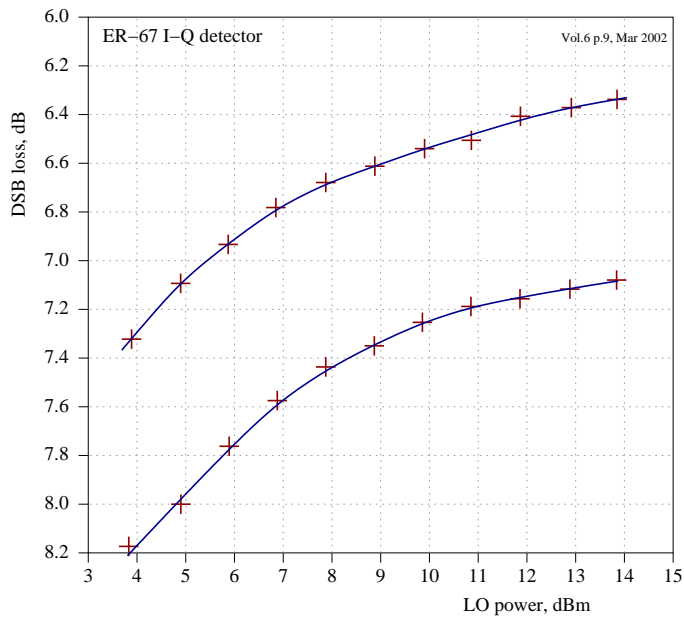}}
\caption{DSB loss of a home-made VHF I-Q detector, based on Mini Circuits mixers and power splitters.}
\label{fig:mix-sd-v6p9fig}
\end{figure}
The Type-2 detector seems to work better than the Type-1 because the $180^\circ$ junction exhibit higher symmetry and lower loss than the $90^\circ$ junction.  Some power loss and asymmetry is more tolerated at the LO port, which is saturated.  Figure~\ref{fig:mix-sd-v6p9fig} gives an idea of actual loss asymmetry.  In addition, there can be a phase error, that is a deviation from quadrature, of a few degrees.

\begin{figure}[t]
\centering%     % COMBINED LATEX-PS/PDF FIGURES
	\centering\scalebox{\smallscale}{%  %              % scale
	\input{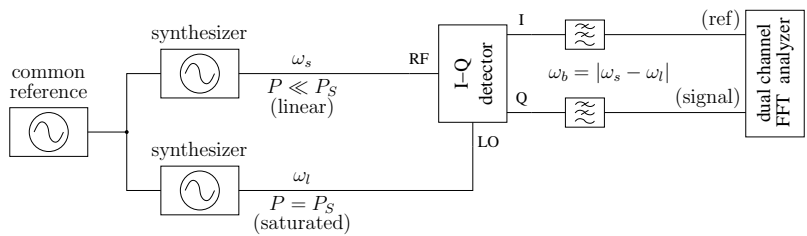}}
\caption{Understanding the phase relationships inside an I-Q detector.}
\label{fig:mix-iq-identify}
\end{figure}
Finally, it is worth pointing out that the phase relationships shown in Figures~\ref{fig:mix-iq-detectors}--\ref{fig:mix-iq-modulators} result from a technical choice, for they should not be given for granted.  Letting the phase of the LO arbitrary, there are two possible choices, Q leads I or Q lags I\@.  The experimentalist may come across unclear or ambiguous documentation,
hence inspection is recommended.  Figure~\ref{fig:mix-iq-identify} shows a possible method.  The FFT analyzer is used to measure the phase of the signal Q versus the reference signal I\@.  
I have some preference for $\omega_s>\omega_l$, and for a beat note $\frac{1}{2\pi}\omega_b=\frac{1}{2\pi}|\omega_s-\omega_b|$ of some 1--5 kHz.  A phase-meter, a vector voltmeter, or a lock-in amplifier can be used instead of the dual-channel FFT analyzer.

%=============================================
\section{Non-ideal behavior}
%=============================================
Most of the issues discussed here resort to the general background on radio-frequency and microwave background, for they are listed quickly only for the sake of completeness. 
The book \cite{razavi:rf-microelectronics} is a good reference.

\begin{figure}[t]
\centering%     % COMBINED LATEX-PS/PDF FIGURES
	\centering\scalebox{\normalscale}{%  %              % scale
	\input{\includefigpathmix-twotone-ip3.pstex_t}}
\caption{.}
\label{fig:mix-twotone-ip3}
\end{figure}

\begin{description}
\item[Impedance matching.]~Inputs and output of the mixer only approximate the nominal impedance, for reflection are present in the circuit.  In practice, the impedance mismatching depends on frequency and power.

\item[Isolation and crosstalk.]~A fraction of the input power leaks to the output, and to the other input as well.  Often, isolating the LO port is relevant because of power. 

\item[1 dB compression point.]  At high input power, of about 10 dB below the LO power, the mixer starts saturating, hence the SSB loss increases.  The 1 dB compression power is defined as the compression power at which the loss increases by 1 dB (Figure~\ref{fig:mix-loss}).

\item[Non-linearity.]~The mixer behavior deviates from the ideal linear model of Section~\ref{ssec:mix:linearity}, for the input-output relationship is of the form 
\begin{math}
v_o(v_i) = a_0 + a_1v_i + a_2v_i^2 + a_3v_i^3 + \ldots~~
\end{math}
[Eq.~\req{eqn:mix-nonlinear-polynomial}, here repeated]. 
In radio engineering the cubic term, $a_3v_i^3$, is often the main concern.  This is due to the fact that, when two strong adjacent-channel signals are present at $\Delta\omega$ and $2\Delta\omega$ off the received frequency $\omega_i$, a conversion product falls exactly at $\omega_i$, which causes interference.  Being $\Delta\omega\ll\omega_i$, a preselector filter can not fix the problem. 

\item[Offset.]~In `synchronous detector' mode, the output differs from the expected value by a dc offset, which depends on the LO power and of frequency.  The same problem is present in the in `phase detector' mode, where also the RF power affects the offset.  This occurs because of saturation. 

\item[Internal phase shift.]~The presence of a small phase lag at each port inside the mixer has no effect in most application.  Of course, in the case of I-Q devices the quadrature accuracy is relevant. 

\end{description}

%==============================================
\section{Mixer Noise}
\begin{figure}[ht]
\centering%     % COMBINED LATEX-PS/PDF FIGURES
	\centering\scalebox{\normalscale}{%  %              % scale
	\input{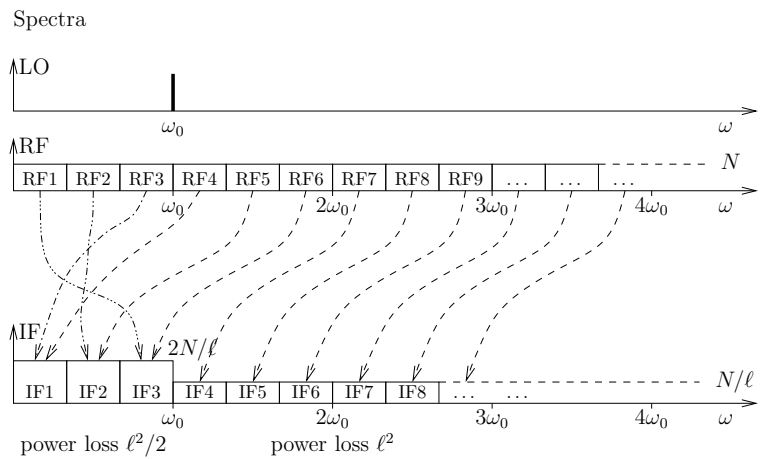}}
\caption{A step appears in the conversion of white noise.}
\label{fig:mix-noise-step}
\end{figure}

%==============================================
The mixer noise were studied since the early time of radars \cite{radlab-v15-torrey:crystal-rectifiers,bergmann68iretmtt}.  Significantly lower noise was later obtained with the Schottky diode \cite{barber67mtt,gewartowski71mtt}, and afterwards with the double balanced mixer.  More recent and complete analysis of the mixer noise is available in \cite{held78mtt-1,held78mtt-2,kerr78mtt-0,kerr78mtt-1,kerr78mtt-2}.  
Nonetheless in the design electronics, and even in \emph{low-noise} electronics, the mixer noise is often a second-order issue because:
\begin{enumerate}  
\item Nowadays mixers exhibit low noise figure, of the order of 1 dB\@.
\item The mixer is almost always preceded by an amplifier.
\item The mixer picks up noise from a number of frequency slots sometimes difficult to predict.
\end{enumerate}
Noise pick-ups from various frequency slots is probably the major practical issue.  The presence of the USB/LSB pair makes the image-rejection mixer (Fig.~\ref{fig:mix-img-rej}, p.~\pageref{fig:mix-img-rej}) appealing.  Two phenomena deserve attention.  The first one is the multi-harmonic frequency conversion (Fig.~\ref{fig:mix-lc-harm} p.~\pageref{fig:mix-lc-harm} and Fig.~\ref{fig:mix-ld-harm} p.~\pageref{fig:mix-ld-harm}), by which noise is converted to the IF band from the sidebands of frequencies multiple of the LO frequency.  The second phenomenon is a step in the output noise spectrum at the LO frequency, in the presence of white noise at the RF port (Fig.~\ref{fig:mix-noise-step}).  Only a graphical proof is given here.  The output slots IF1, IF2, and IF3 are down-converted from the input slots RF3+RF4, RF2+RF5, and RF1+RF6, respectively.  Thus, the conversion power loss is $\ell^2/2$.  At higher frequencies, the output slots IF4, IF5, \ldots, come from RF7, RF8, \ldots, for the  loss is $\ell^2$.  The analytical proof follows exactly the graphical proof, after increasing to infinity the number of frequency slots  so that their width is $d\omega$.

Flicker ($1/f$) noise is generally not documented.  All the references found about the mixer noise are limited to classical white noise, that is, thermal and shot noise, while the flicker noise is not considered.  The flicker behavior of mixer may depend on the operating mode, as listed in Table~\ref{tab:mix:modes} (p.~\pageref{tab:mix:modes}).  Yet, the general rule is that flicker noise is a near-dc phenomenon, powered by the LO pump.  Then, the near-dc flicker is up-converted by non-linearity and brougt to the output; or available at the output, in the `synchronous detector' mode (Sec.~\ref{ssec:mix:sd-mode}) and in the `phase detector' mode (Sec.~\ref{sec:mix:pd-mode}), where the dc signal is taken at the output.

%================================================================
\section{Where to learn more}
%================================================================
%
Our approach, which consists of identifying and analyzing the modes of
Table~\ref{tab:mix:modes}, is original.  Thus, there are no specific
references.

A lot can be learned from the data sheets of commercial mixers and
from the accompaining application notes.  Unfortunately, learning in
this way requires patience because manufacturer tend to use their own
notation, and because of the commercial-oriented approach.  Another
problem is that the analysis is often too simplified, which makes
difficult to fit technical information into theory.  Watkins Johnson\footnote{http://www.wj.com/technotes/}
application notes \cite{wj:mixers-1,wj:mixers-2} provide useful
general description and invaluable understanding of intermodulation
\cite{wj:selecting-mixers}.  We also found useful the Anzac
\cite{anzac:mixers,anzac:modulators}, Macom \cite{macom:mixers} and
Mini-Circuits
\cite{minicircuits:understanding-mixers,minicircuits:mixer-terms}
application notes.

Reading books and book chapters on mixers, one may surprised by the
difference between standpoints.  A book edited by E. L.
Kollberg~\cite{kollberg:mixers} collects a series of articles, most of
which published in the IEEE Transactions on Microwave Theory and
Technology and other IEEE Journals.  This collection covers virtually all relevant topics.
The non-specialist may be interested at least in the first part, about basic mixer theory.  The classical book written by S. A. Maas~\cite{maas:mixers} is a must on the subject.  

A few books about radio engineering contains a chapter on mixers.  We
found useful chapter 3 (\emph{mixers}) of McClaning \&
al.~\cite[pp.~261--344]{mcclaning:receivers}, chapter~7
(\emph{Mixers}) of Krauss \&
al.~\cite[pp.~188--220]{krauss:radio-engineering}, chapter~6
(\emph{Mixers}) of Rohde \&
al.~\cite[pp.~277--318]{rohde:communications-receivers}, and Chapter 7
(\emph{Microwave Mixer Design}), of Vendelin \&
al.\cite{vendelin:microwave-circuit-design}.  

Some radio amateur handbooks provide experiment-oriented information of great value, hard
to find elsewere.  Transmission-line transformers and baluns are
described in \cite{sevick:transmission-line}.  Recent editions of the
the ARRL Handbook~\cite{straw:arrl-handbook-99} contain a chapter on
mixers (chapter 15 in the 1999 edition), written by
D.  Newkirk and R.  Karlquist, full of practical information and
common sense.

%==================================
\def\bibfile#1{#1}
%\def\bibfile#1{/home/rubiola/docs/bib/#1}
%==================================
\addcontentsline{toc}{section}{References}
\bibliographystyle{amsalpha}
\bibliography{\bibfile{ref-short},%
              \bibfile{references},%
              \bibfile{rubiola}}

\end{document}